\documentclass[review,5p,times,twocolumn,nopreprintline]{elsarticle}
\usepackage{lineno,hyperref}
\modulolinenumbers[5]

\usepackage{amsthm,amsfonts,amsmath,amssymb}
\usepackage{float}
\usepackage{bm}
\usepackage{verbatim}
\usepackage{overpic}
\usepackage{appendix}

\newcommand{\mb}[1]{\bm{#1}}
\newcommand{\pd}{\partial}
\DeclareMathOperator*{\argmin}{arg\,min}

\graphicspath{{../fig/}{fig/}}

\usepackage{epstopdf}
\AppendGraphicsExtensions{.tif}

\newcommand{\key}[1]%
{\raisebox{0.3mm}{\protect\includegraphics{key/l/#1}}}

\journal{International Journal of Multiphase Flow}

\usepackage{xcolor}
\newcommand{\rev}[1]{{#1}}

\definecolor{crimson}{HTML}{c90016}
\definecolor{forestgreen}{rgb}{0.13, 0.55, 0.13}

\newcommand{\revv}[1]{{#1}}
\newcommand{\revvg}[1]{{#1}}

\begin{document}

\begin{frontmatter}

  \title{A hybrid particle volume-of-fluid method for curvature estimation in multiphase flows\tnoteref{doi}}

  \makeatletter
  \tnotetext[doi]{Published in \@journal\vspace{-4pt}\\\hspace*{14.5pt}\url{https://doi.org/10.1016/j.ijmultiphaseflow.2020.103209}}
  \makeatother

\author{Petr Karnakov}
\author{Sergey Litvinov}
\author{Petros Koumoutsakos\corref{mycorr}}
\ead{petros@ethz.ch}
  \address{
    Computational Science and Engineering Laboratory,
    ETH Z{\"u}rich, Clausiusstrasse 33, CH-8092 Z{\"u}rich, Switzerland}
\cortext[mycorr]{Corresponding author}

\begin{abstract}
\rev{
We present a particle method for estimating the curvature of interfaces  in volume-of-fluid simulations of multiphase flows.  The method is well suited for under-resolved interfaces, and it is shown to be more accurate than the parabolic fitting that is employed in such cases. The curvature is computed from the equilibrium positions of particles constrained to circular arcs and attracted to the interface. The proposed particle method is combined with the method of height functions at higher resolutions, and it is shown to outperform the current combinations of height functions and parabolic fitting.  The algorithm is conceptually simple and straightforward to implement on new and existing software frameworks for multiphase flow simulations thus enhancing their capabilities in challenging flow problems. We evaluate the proposed hybrid method on a number of two- and three-dimensional benchmark flow problems and illustrate its capabilities on simulations of flows involving bubble  coalescence and turbulent multiphase flows. 
}
\end{abstract}


\end{frontmatter}

\section{Introduction}

Bubbles and drops are critical components of important industrial
applications such as boiling and condensation~\cite{kharangate2017},
bubble column reactors~\cite{jakobsen2005}, electrochemical cells
\cite{boissonneau2000} and physical systems involving air entrainment in
plunging jets~\cite{kiger2012} and liquid jet atomization~\cite{ling2015}.
Simulations of such processes are challenged by the multiple scales of
bubbles and their surface tension.  Since the pioneering work of Brackbill
\textit{et al.}~\cite{brackbill1992} in modeling surface tension
with the Eulerian representation, 
a number of advances have been made~\cite{popinet2018},
using level-sets~\cite{sussman1998} and volume-of-fluid methodologies
(VOF)~\cite{scardovelli1999} to describe the interface and  compute the
surface tension.  

The reconstruction of the interface in VOF methods is
prone to inaccuracies  that were shown to be eliminated in certain cases
through a parabolic reconstruction of interfaces~\cite{renardy2002}
and the balancing of pressure gradients with surface tension.
The interface curvature estimation was further improved by the method
of height functions~\cite{cummins2005} that employs the discrete volume
fraction field.  The algorithm chooses a coordinate plane and integrates
the volume fraction in columns perpendicular to the plane to obtain a
function representing the distance from the interface  to the plane. A
well-defined height corresponds to a column crossing the interface exactly
once such that its endpoints are on the opposite sides of the interface.
The curvature is then estimated by finite differences on the plane which
allows for high-order convergence~\cite{sussman2006}.  However, the
method requires that the heights are available on a sufficiently large
stencil which imposes strong restrictions on the resolution: five cells
per radius for circles and eight cells for spheres~\cite{popinet2009}.
Modifications of the method aim to weaken this requirement by fitting an
analytical function to the known values~\cite{renardy2002, bornia2011,
diwakar2009}.  Heuristic criteria define whether the method of height
functions or its modifications are applied in every cell.  The first
complete implementation of such approach was given by the generalized
height-function (GHF) method~\cite{popinet2009} which used parabolic fitting
to heights from mixed directions and to centroids of the interface
fragments.  A similar approach was later implemented by~\cite{ling2015}.
An alternative approach is the mesh-decoupled height function method
\cite{owkes2015} allowing for arbitrary orientation of the columns.
Each height is computed from the intersection of the column, and the fluid
volume reconstructed by polyhedrons.  However, in three dimensions the
procedure involves complex and computationally expensive geometrical
routines for triangulation of the shapes and still requires at least
three cells per radius.  The method of parabolic reconstruction directly
from the volume fractions~\cite{evrard2017} has a high-order convergence
rate without restrictions on the minimal resolution in two dimensions.
However, the extension of this algorithm to three dimensions is not
straightforward.

We introduce a new method for computing the curvature in the
volume-of-fluid framework which allows for solving transport problems
with bubbles and drops at low resolution up to one cell per radius.
The method relies on a reconstruction of the interface,
and it is applicable, but not limited, to the volume-of-fluid methods.
The curvature estimation is obtained by fitting circular arcs to the
reconstructed interface.  A circular arc is represented by a string of
particles.  The fitting implies an evolution of the particles
under constraints with forces attracting them to the interface.

We remark that the present approach is related to the concept of active
contours~\cite{kass1988}. The key differences include the imposition
of hard constraints (particles belong to circular arcs) and the use
of attraction forces based on the interface reconstruction.
\revv{Our approach is also different from the finite particle
method~\cite{wenzel2018} which uses particles to construct a smooth
representation of the volume fraction field: the particles are assigned with
weights computed by averaging the volume fraction over neighboring cells,
but their positions remain constant.
On the other hand, out method only uses the positions of particles
determined through an equilibration process.}
We note that the present algorithm 
is more accurate than
the generalized height-function method~\cite{popinet2009,ling2015} 
up to a resolution of four cells per curvature radius
in three dimensions and even at a resolution of one cell per radius 
provides the relative curvature error below 10\%.

The paper is organized as follows. Section~\ref{s:meth} describes the
method for curvature estimation and the model of flows with
surface tension as an application.  Section~\ref{s:test} reports results
on test cases involving spherical interfaces.  Section~\ref{s:app}
presents applications to turbulent flows and bubble coalescence.
Section~\ref{s:concl} concludes the study.


\section{Numerical methods}
\label{s:meth}
In this section,
we describe a standard numerical model
for two-component incompressible flows with surface tension
and introduce our particle method for estimating the interface curvature.

\subsection{VOF method for multiphase flows with surface tension}
\label{s:flow}

\revv{We consider the numerical model describing
two-component incompressible flows with surface tension
available in the open-source solver Basilisk~\cite{basilisk,popinet2009}.}
The system consists of the Navier-Stokes equations 
for the mixture velocity~$\mb u$ and pressure~$p$
\begin{align}
  \nabla \cdot \mb u &= 0, \\
  \rho\Big(\frac{\pd \mb u}{\pd t} + (\mb u \cdot \nabla)\, \mb u\Big)
  &=
  -\nabla p + \nabla \cdot \mu (\nabla \mb u + \nabla \mb u^T )
  + \mb f_\sigma + \rho\mb g
\end{align}
and the advection equation for the volume fraction~$\alpha$
\begin{equation}
  \frac{\pd \alpha}{\pd t} + (\mb u \cdot \nabla)\, \alpha = 0
\end{equation}
with
density~$\rho=(1-\alpha)\rho_1 + \alpha\rho_2$,
dynamic viscosity~$\mu=(1-\alpha)\mu_1 + \alpha\mu_2$,
gravitational acceleration~$\mb g$
and constant material parameters 
$\rho_1$, $\rho_2$, $\mu_1$ and $\mu_2$.
The surface tension force is defined as
$\mb f_\sigma = \sigma \kappa \nabla \alpha$
with the surface tension coefficient $\sigma$
and interface curvature $\kappa$.
A finite volume discretization is based on
Chorin's projection method for the pressure coupling~\cite{chorin1968}
and the Bell-Colella-Glaz scheme~\cite{bell1989} for convective fluxes.
The advection equation is solved using the volume-of-fluid method PLIC
with piecewise linear reconstruction~\cite{weymouth2010}
where the normals are computed using the mixed Youngs-centered scheme
which is a combination of Youngs' scheme and the height functions.
The approximation of the surface tension force is 
well-balanced~\cite{francois2006}
(i.e. the surface tension force is balanced by the pressure gradient
if the curvature is uniform)
and requires face-centered values of the curvature,
which are computed as the average over the neighboring cells.
We refer to \cite{aniszewski2019} for further details
about the algorithm.

\revv{To compute the cell-centered interface curvature,
we use the proposed algorithm described in the following sections.
Our implementation of the curvature estimator is available online}%
\footnote{Implementation with Basilisk:
\href{https://cselab.github.io/aphros/basilisk_partstr.zip}
{https://cselab.github.io/aphros/basilisk\_partstr.zip}}
and can be directly used in Basilisk for simulations
of multiphase flows with surface tension.
We also provide a visual web-based demonstration of the method%
\footnote{Visual demonstration:
\href{https://cselab.github.io/aphros/curv.html}
{https://cselab.github.io/aphros/curv.html}}
and a reference implementation in Python%
\footnote{Implementation in Python: 
\href{https://cselab.github.io/aphros/curv.py}
{https://cselab.github.io/aphros/curv.py}}.

\subsection{\revv{Particles for estimating the curvature from line segments}}
\label{s:meth2d}

This section introduces a particle method
for estimating the curvature of an interface
represented by a set of line segments,
which are given by the piecewise linear reconstruction
of the interface as described in the next section.
The particles are constrained to circular arcs and equilibrate
on the interface due to attraction forces.

Consider a set of $N_L$ line segments $[\mb{a}_l, \mb{b}_l],\;l=1,\dots,N_L$
with endpoints~$\mb{a}_l,\mb{b}_l\in\mathbb{R}^2$.
One is distinguished as the target line segment $[\mb{a}^*, \mb{b}^*]$
at which the curvature is to be estimated.
The union of all line segments is denoted as
\begin{equation}
  L=\bigcup_{l=1}^{N_L}\,[\mb{a}_l, \mb{b}_l].
\end{equation}

\begin{figure*}
  \centering
  \raisebox{4.5cm}{(a)}
  \includegraphics[scale=0.75]{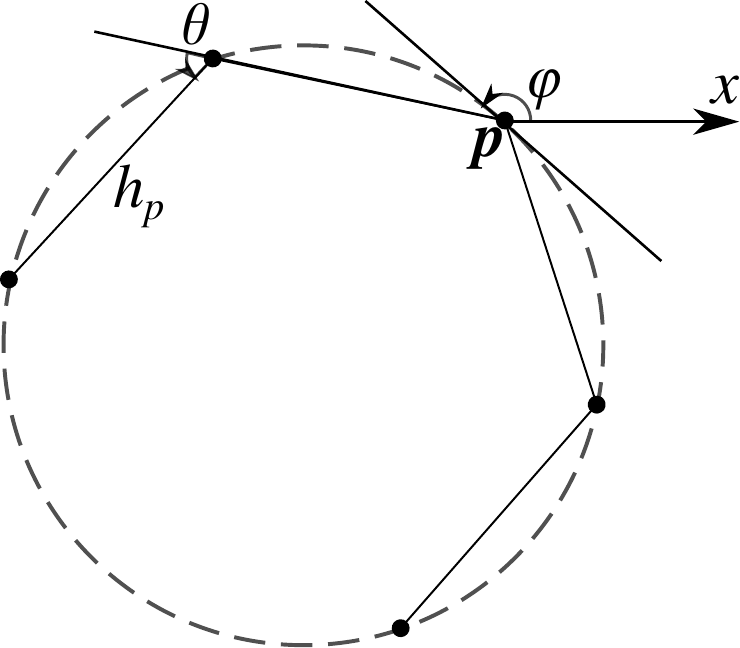}
  \hspace{1cm}
  \raisebox{4.5cm}{(b)}
  \includegraphics[scale=0.75]{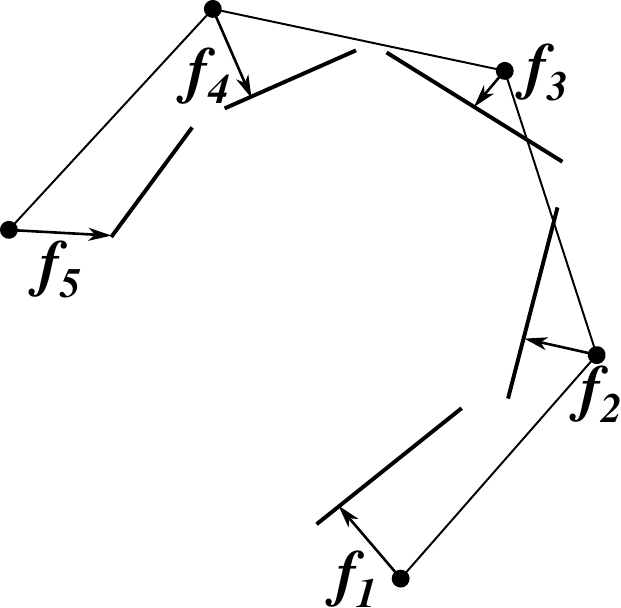}

  \caption{%
    (a) All particles belong to a circle,
    and their positions are defined by the central particle~$\mb{p}$, 
    the orientation angle~$\phi$ and the bending angle~$\theta$.
    (b) Line segments represent the interface,
    and the force acts on each particle towards the nearest 
    point on the interface.
  }
  \label{f:sk_constr}
\end{figure*}

Given an odd number $N$,
we introduce a string of $N$ particles
$\mb{x}_i\in\mathbb{R}^2,\; i=1,\dots,N$
and denote the index of the central particle as $c=(N+1)/2$.
The particles are constrained to circular arcs,
which leads to parametrization of their positions
\begin{equation}
  \label{e:partpos}
  \mb{x}_i(\mb{p},\phi,\theta) =
  \begin{cases}
  \mb{p} + \sum\limits_{j=1}^{i-c} h_p
    \mb{e}\big(\phi + (j-\frac{1}{2})\,\theta\big) &\qquad i> c,
    \\
  \mb{p} &\qquad i=c,
    \\
  \mb{p} - \sum\limits_{j=1}^{c-i} h_p
    \mb{e}\big(\phi - (j-\frac{1}{2})\,\theta\big) &\qquad i<c,
  \end{cases}
\end{equation}
where $\mb{p}$~is the origin,
$\phi$~is the rotation angle, $\theta$~is the bending angle,
$h_p = {H_p h}/(N-1)$ is the distance between neighboring particles
and $\mb{e}(\psi)=\cos{\psi}\,\mb{e}_x + \sin{\psi}\,\mb{e}_y$.
Parameter $H_p$ defines the length of the string relative to the mesh step~$h$.
Values of~$N$ and $H_p$ are chosen in Section~\ref{s:sens}.
The curvature of the circular arc is related to the bending angle as
\begin{equation}
  \label{e:kap}
  \kappa(\theta)=\frac{2\sin\tfrac{\theta}{2}}{h_p}.
\end{equation}
We define the force attracting a particle at position $\mb{x}$
to the nearest point on the interface
\begin{gather}
  \label{e:force}
  \mb{f}(\mb{x}) = \eta\, (\argmin_{\mb{y}\in L}|\mb{y}-\mb{x}| - \mb{x}),
\end{gather}
where $\eta\in [0,\,1]$ is a relaxation factor.
Figure~\ref{f:sk_constr} illustrates the parametrisation of positions
and computation of forces.

Further derivations use vector notation of the form
$\mb{X} = [\mb{x}_i,\;i=1,\dots,N]$,
where $\mb{x}_i$ is component $i$ of $\mb{X}$.
The scalar product is defined as
$\mb{X}\cdot\mb{Y}=\sum\limits_{i=1}^{N} \mb{x}_i \cdot \mb{y}_i$.
In this notation, the positions and forces combine to
\begin{align}
  \mb{X}(\mb{p}, \phi, \theta) 
  &= 
  [\mb{x}_i(\mb{p}, \phi, \theta),\;i=1,\dots,N],
  \\
  \label{e:forcevect}
  \mb{F}(\mb{p}, \phi, \theta) 
  &= 
  [\mb{f}(\mb{x}_i(\mb{p}, \phi, \theta)),\;i=1,\dots,N].
\end{align}

The particles evolve until equilibration according to the following
iterative procedure.
Initially, the particles are arranged along
the target line segment~$[\mb{a}^*,\mb{b}^*]$.
The central particle is placed at the segment center
$\mb{p}^0=(\mb{a}^*+\mb{b}^*)/2$,
the rotation angle~$\phi^0$ is given by vector $\mb{b}^*-\mb{a}^*$,
and the bending angle is zero $\theta^0=0$.
Iteration~$m+1$ consists of three steps, each correcting one parameter:
\paragraph{Step 1} Compute positions and forces, correct~$\mb{p}$
by the force on the central particle
and subtract the change of positions from the forces
\begin{gather*}
  \mb{X}^m = \mb{X}(\mb{p}^m, \phi^m, \theta^m),\quad
  \mb{F}^m = \mb{F}(\mb{p}^m, \phi^m, \theta^m) \\
  \mb{p}^{m+1} = \mb{p}^m + \mb{f}_c^m \\
  \mb{X}^* = \mb{X}(\mb{p}^{m+1}, \phi^m, \theta^m),\quad
  \mb{F}^* = \mb{F}^m - (\mb{X}^* - \mb{X}^m).
\end{gather*}
\paragraph{Step 2} Correct $\phi$ by projection of force on derivative 
and subtract the change of positions from the forces
\begin{gather*}
  \mb{D}_\phi = \frac{\pd \mb{X}}{\pd \phi}(\mb{p}^{m+1}, \phi^m, \theta^m)\\
  \phi^{m+1} = \phi^m +
    \frac{\mb{F}^*\cdot\mb{D}_\phi}{\mb{D}_\phi\cdot\mb{D}_\phi}  \\
  \mb{X}^{**} = \mb{X}(\mb{p}^{m+1}, \phi^{m+1}, \theta^m),\quad
  \mb{F}^{**} = \mb{F}^* - (\mb{X}^{**} - \mb{X}^*).
\end{gather*}
\paragraph{Step 3} Correct $\theta$ by projection of force on derivative
\begin{gather*}
  \mb{D}_\theta = 
    \frac{\pd \mb{X}}{\pd \theta}(\mb{p}^{m+1}, \phi^{m+1}, \theta^m) \\
  \theta^{m+1} = \theta^m +
    \frac{\mb{F}^{**}\cdot\mb{D}_\theta}{\mb{D}_\theta\cdot\mb{D}_\theta}.  \\
\end{gather*}
Expressions for derivatives
$\frac{\pd\mb{X}}{\pd\phi}$ and $\frac{\pd\mb{X}}{\pd\theta}$
are provided in Section~\ref{s:partderiv},
and a proof of convergence is given in Section~\ref{s:proof}.

The iterations are repeated until
\begin{equation}
  E_m = \frac{\max_i |\mb{x}_i^{m} - \mb{x}_i^{m-1}|_\infty}{\eta\, h}
  < \varepsilon_p
  \quad\text{or}\quad m > m_\text{max},
  \label{e:eta}
\end{equation}
where $E_m$ is the maximum difference after iteration $m$,
$h$ is the mesh step,
$\eta$ is the relaxation parameter from (\ref{e:force}),
$\varepsilon_p$ is the convergence tolerance and 
$m_\text{max}$ is the maximum number of iterations.
Values for these parameters are chosen in Section~\ref{s:conviter}.
Finally, the curvature is computed from the bending angle~$\theta$
using relation~(\ref{e:kap}).

Steps~2 and~3 are defined to provide the correction of positions
which minimizes the distance to forces.
With linear approximation in terms of ~$\phi^{m+1}-\phi^m$, 
this leads to a minimization problem
\begin{equation}
  \big\|\mb{F}^*-
  (\phi^{m+1}-\phi^{m})\mb{D}_\phi
  \big\|_2 
  \rightarrow \text{min}
\end{equation}
and gives the optimal correction
\begin{equation}
  \phi^{m+1} =  \phi^m +
    \frac{\mb{F}^*\cdot\mb{D}_\phi}
    {\mb{D}_\phi \cdot \mb{D}_\phi}.
\end{equation}
The same procedure is applied to the bending angle~$\theta$.
We note that using the same principle for Step~1
would correspond to correcting the origin~$\mb{p}$
by the mean force~$\frac{1}{N}\sum_{i=1}^N\mb{f}_i^m$ instead of~$\mb{f}_c^m$.
However, simulations on the test case with a
static droplet introduced in Section~\ref{s:static}
showed that this would result in stronger spurious flows
and lack of equilibration.

We also observe that, for the chosen initial conditions and forces,
Step~1 trivializes to $\mb{p}^m=\mb{p}^0$
since the origin already belongs to a line segment,
which results in zero force $\mb{f}_c$.
However, we still include this step in the algorithm
to allow for modifications of the attraction force.
One such modification, described in Section~\ref{s:circ},
consists in replacing the line segments
with circular arcs and recovers the exact curvature
if the endpoints of all line segments belong to a circle.

\subsection{\revv{Particles for estimating the curvature from volume fraction field}}
\label{s:curv3d}

We estimate the interface curvature from a discrete volume fraction field
defined on a uniform Cartesian mesh.
Our method uses a piecewise linear reconstruction of the interface,
which is available as part of the PLIC technique
for solving the advection equation.
Following~\cite{aulisa2007}, we compute the interface
normals with the mixed Youngs-centered method
on a $3\times3\times3$ stencil in 3D (or $3\times3$ in 2D).
To define the orientation, we assume that
the normals have the direction of anti-gradient of the volume fraction.
Then, the interface is reconstructed in each cell independently by a
polygon (or line segment)
cutting the cell into two parts according to the estimated
normal and the given volume fraction~\cite{scardovelli2000}.

In two dimensions, the interface is represented as a set of line segments.
To estimate the curvature in one interfacial cell, we collect
the line segments from a~$5\times5$ stencil centered at the target cell.
The stencil of this size includes all line segments
that can be reached by the particles if the string length is $H_p\leq4$.
We apply the algorithm from Section~\ref{s:meth2d}
to these line segments and compute the curvature.

\begin{figure*}
  \centering
  \raisebox{3cm}{(a)}
  \includegraphics[scale=0.75]{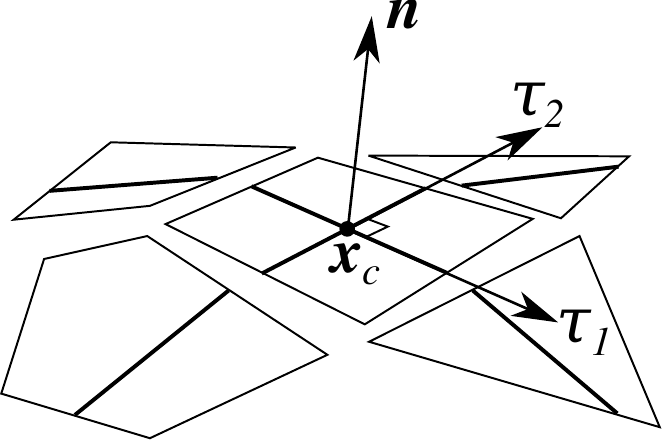}
  \hspace{2cm}
  \raisebox{3cm}{(b)}
  \raisebox{0.5cm}{\includegraphics[scale=0.75]{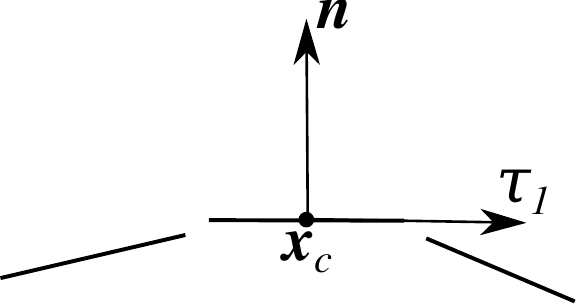}}

  \caption{%
    Curvature estimation in three dimensions in case $N_s=2$.
    (a) Interface polygons with the cross sections.
    (b) One cross section in local two-dimensional coordinates.
  }
  \label{f:sk_3d}
\end{figure*}

In three dimensions, the reconstructed interface is a set 
of planar convex polygons.
In this case, we compute the mean curvature as the average
over $N_s$ cross sections and thus reduce
the problem to the two-dimensional case.
The algorithm to estimate the mean curvature in a cell
consists of the following steps:
\paragraph{Step 1}
Collect a set~$P$ of polygons from a~$5\times5\times5$ stencil
centered at the target cell.
Determine the unit normal~$\mb{n}$ of the target polygon
and the center~$\mb{x}_c$ as the mean over its vertices.
\paragraph{Step 2}
Compute the curvature~$\kappa_j$  
in each cross section $j=0,\dots,N_s-1$:
\begin{itemize}
  \item 
    Define a plane passing through $\mb{x}_c$ and containing vectors~$\mb{n}$ 
    and 
    \begin{equation*}
    \mb{\tau}=\cos(\pi j / N_s)\mb{\tau}_1 + \sin(\pi j / N_s)\mb{\tau}_2,
    \end{equation*}
    where 
    $\mb{\tau}_1 = \mb{n} \times \mb{e}$,
    $\mb{\tau}_2 = \mb{n} \times \mb{\tau}_1$
    and $\mb{e}$ is one of the unit vectors $\{\mb{e}_x, \mb{e}_y, \mb{e}_z\}$
    providing the minimal~$|\mb{n}\cdot\mb{e}|$.
  \item Intersect each polygon in $P$ with the plane
    and collect non-empty intersections in a set of line segments
    $\{[\mb{a}_l,\mb{b}_l]\}$.
  \item Construct a set 
    $\{[\hat{\mb{a}}_l, \hat{\mb{b}}_l]\}$
    from
    $\{[\mb{a}_l,\mb{b}_l]\}$
    by computing local two-dimensional coordinates of the endpoints
    \begin{equation*}
      \hat{\mb{x}}= 
    \big((\mb{x} - \mb{x}_c) \cdot \mb{\tau}, 
         (\mb{x} - \mb{x}_c) \cdot \mb{n} \big) \in\mathbb{R}^2,
    \end{equation*}
    where $\mb{x}=\mb{a}_l$ or $\mb{x}=\mb{b}_l$.
  \item Apply the procedure from Section~\ref{s:meth2d} to
    $\{[\hat{\mb{a}}_l, \hat{\mb{b}}_l]\}$
    and compute the curvature $\kappa_j$.
\end{itemize}
\paragraph{Step 3}
Compute the mean curvature
\begin{equation}
  \kappa = \frac{1}{N_s} \sum_{j=0}^{N_s-1} {\kappa_j}.
\end{equation}
Figure~\ref{f:sk_3d} illustrates the algorithm.
This approach of cross-sections allows us to estimate the mean curvature
in three dimensions
by using the particle method formulated for line segments on a plane.

\subsection{Convergence of iterations}
\label{s:conviter}
\begin{figure}
  \centering
  \includegraphics{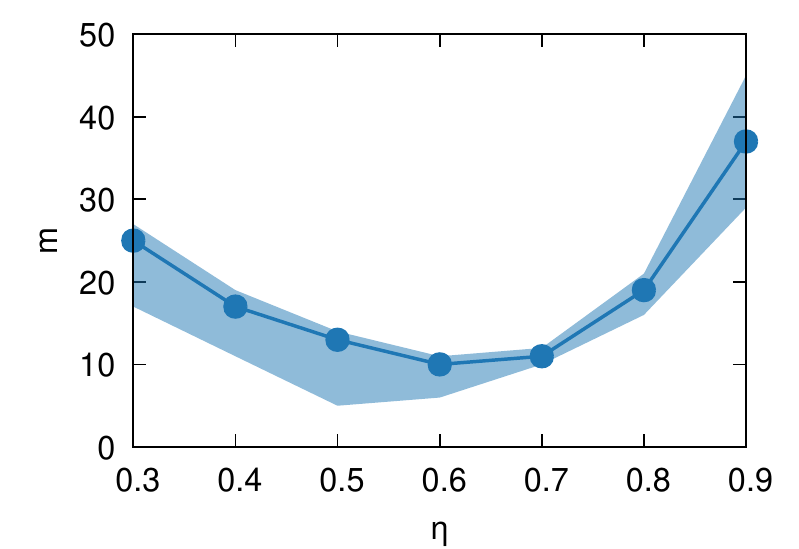}
  \includegraphics{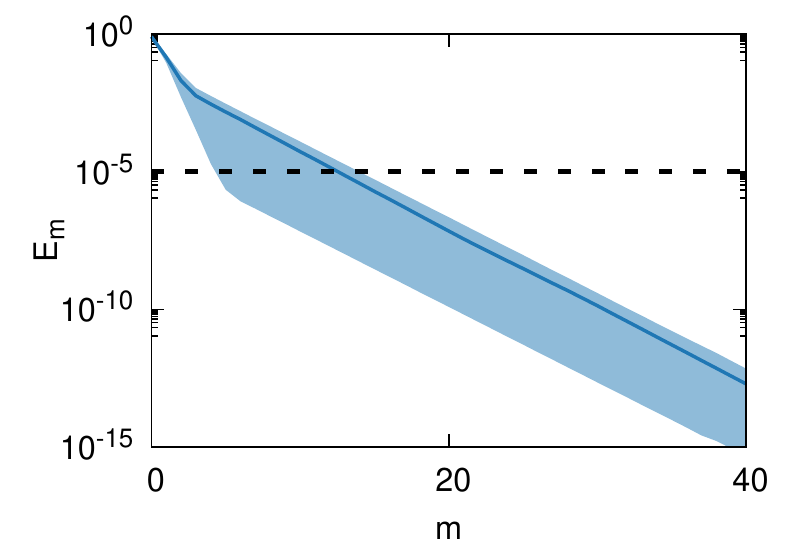}
  \caption{%
    Equilibration of particles on a sphere at resolution $R/h=4$.
    The number of iterations required to reach
    the convergence criterion~(\ref{e:eta})
    with~$\varepsilon_p=10^{-5}$ 
    depending on the relaxation factor~$\eta$ (top)
    and the convergence history for the difference after
    one iteration with~$\eta=0.5$ (bottom).
    The lines show the median
    and the shades show the 10\% and 90\% percentiles 
    over the interfacial cells.
  }
  \label{f:iter}
\end{figure}

\revv{The iterative algorithm in Section~\ref{s:meth2d}
is guaranteed to converge for
a sufficiently small relaxation factor as shown in Section~\ref{s:proof}.}
Here we demonstrate the convergence on the test case of
estimating the curvature of a sphere introduced in Section~\ref{s:curv}.
The convergence tolerance is set to $\varepsilon_p=10^{-5}$
which is three orders of magnitude smaller than the typical curvature error.
Figure~\ref{f:iter} shows
the number of iterations required to satisfy
the convergence criterion~(\ref{e:eta}) depending on the relaxation factor.
We choose the relaxation factor of $\eta=0.5$
for which the equilibration takes 10-20 iterations
as seen from the convergence history in Figure~\ref{f:iter}.
All further computations are performed with
$\eta=0.5$, $\varepsilon_p=10^{-5}$ and $m_\text{max}=20$.

\subsection{\rev{Generalized height-function method}}

We compare our method with the generalized height-function method
for curvature estimation implemented in Basilisk%
\footnote{We use the version of Basilisk available as of August 21, 2019}
~\cite{basilisk,popinet2009}.
The algorithm attempts to compute the curvature in each interfacial cell with 
a series of techniques depending on the resolution.
In three dimensions they can be outlined as follows
\begin{itemize}
  \item Evaluate the height function along a preferred coordinate plane
    chosen based on the interface normal.
    The heights are obtained by summation over columns that span up to 13 cells.
    If consistent heights are available on the $3\times 3$ stencil,
    compute the curvature from them with finite differences.
  \item Collect heights from mixed directions in the $3\times 3\times 3$
    stencil. If found six consistent heights,
    compute the curvature from a paraboloid fitted to them.
  \item If the curvature is already defined in one of the neighboring cells 
    in the $3\times3\times3$ stencil, copy the curvature from the neighbor.
\item Collect centroids of interfacial cells in the $3\times3\times3$ stencil.
  If found six centroids,
  compute the curvature from a paraboloid fitted to them.
\item If all techniques fail, set the curvature to zero.
\end{itemize}

We refer to the documentation of
Basilisk%
\footnote{Curvature estimation in Basilisk:
\href{http://basilisk.fr/src/curvature.h}
{http://basilisk.fr/src/curvature.h}}
for further details of the algorithm.
As clearly seen from the above description,
the generalized height-function method,
while providing an algorithm applicable at all resolutions,
involves four different sources of curvature:
heights, parabolic fitting to heights from mixed directions,
values of curvature from neighboring cells and 
parabolic fit to centroids of the piecewise linear interface.
Such procedure is complex for implementation and lacks robustness
(e.g. depends on the presence of
neighboring interfacial cells for parabolic fitting).

\subsection{\rev{Combined particle and height-function method}}
\label{s:comb}

As shown in Section~\ref{s:test},
the error of our method asymptotically approaches a constant.
To achieve second-order convergence,
we follow the idea of the generalized height-function method
and switch to standard heights at high resolutions.
Our combined method therefore consists of two steps
\begin{itemize}
  \item Evaluate the height function along a preferred coordinate plane
    chosen based on the interface normal.
    If consistent heights are available on a $3\times 3$ stencil,
    compute the curvature from them with finite differences.
  \item Otherwise, compute the curvature using the proposed particle
    method as described in Section~\ref{s:curv3d}.
\end{itemize}

Overall, the logic of the algorithm has greatly simplified
compared to the generalized height function method.
At the same time, 
as we demonstrate in the following sections,
the combined method provides better accuracy at low resolutions
and second-order convergence for well-resolved interfaces.

\subsection{Sensitivity to parameters}
\label{s:sens}

\begin{figure*}
  \centering
  \includegraphics{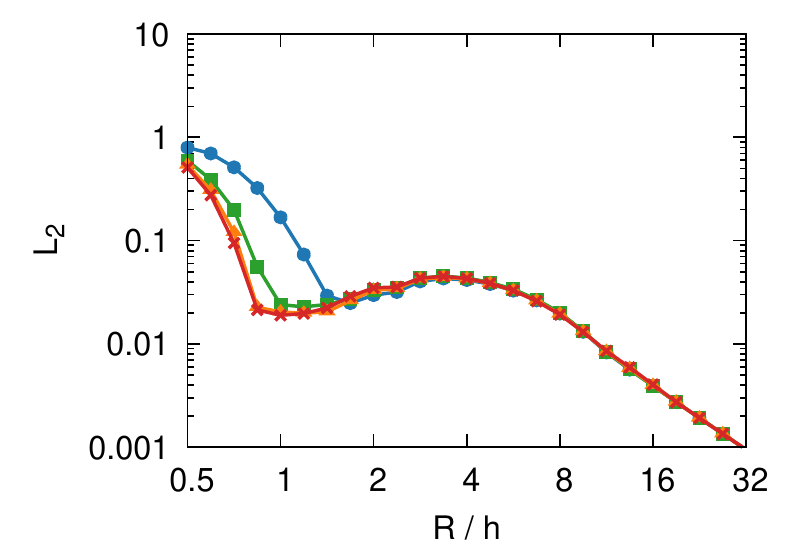}%
  \includegraphics{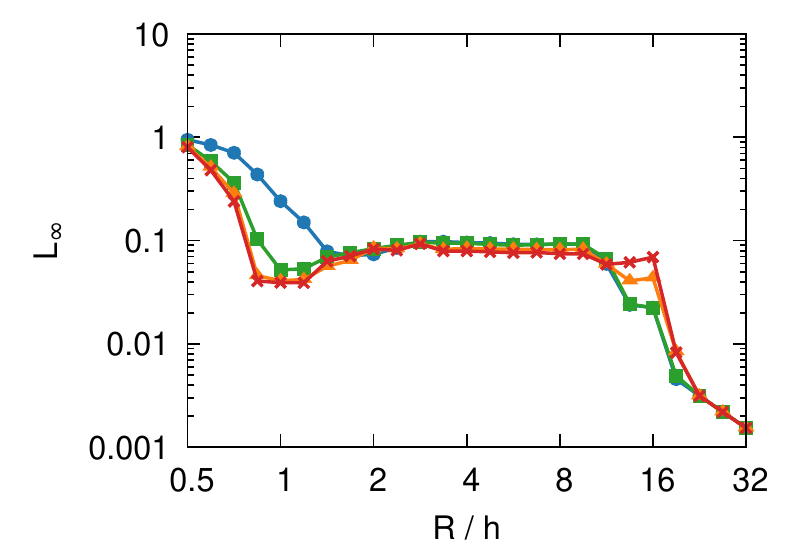}
  \caption{%
    Curvature error for a sphere in $L_2$ and $L_\infty$ norms
    depending on the resolution
    for various values of the number of particles
    $N=3$~\key{011}, 
    $5$~\key{032}, 
    $7$~\key{023} 
    and $9$~\key{044}.
  }
  \label{f:varynp}
\end{figure*}

\begin{figure*}
  \centering
  \includegraphics{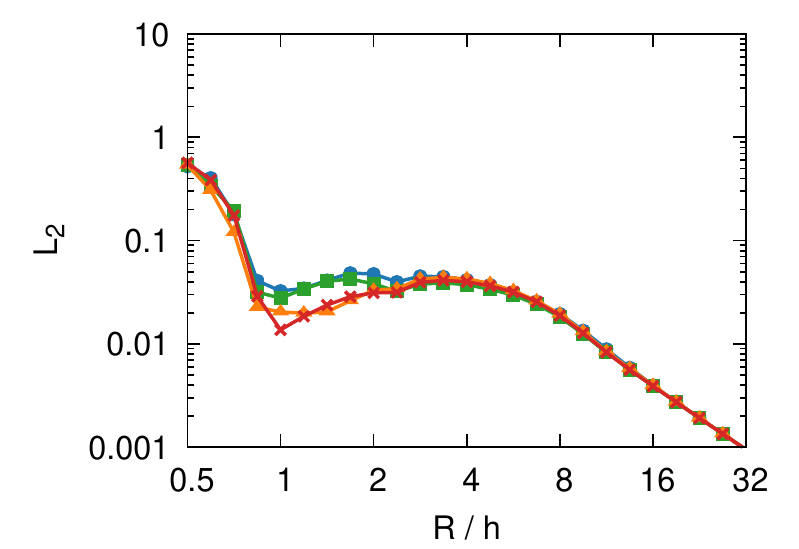}%
  \includegraphics{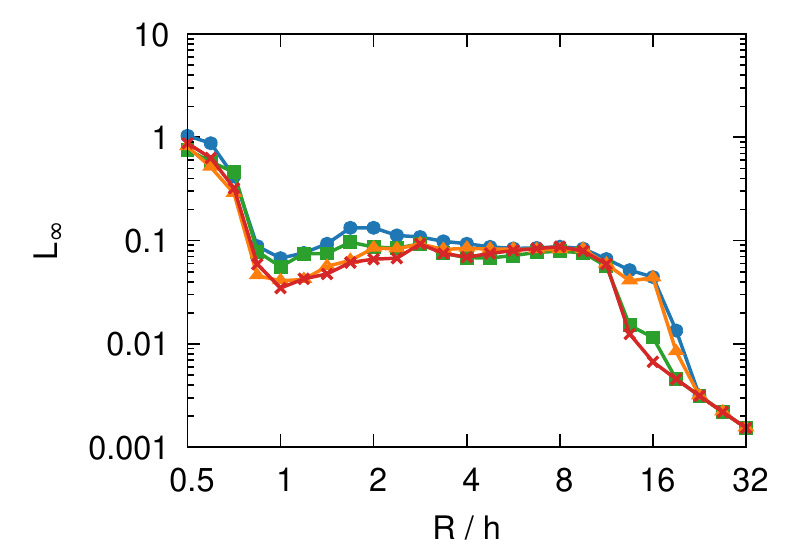}
  \caption{%
    Curvature error for a sphere in $L_2$ and $L_\infty$ norms
    depending on the resolution
    for various values of the number of cross sections
    $N_s=1$~\key{011}, 
    $2$~\key{032},
    $3$~\key{023}
    and $4$~\key{044}.
  }
  \label{f:varyns}
\end{figure*}

\begin{figure*}
  \centering
  \includegraphics{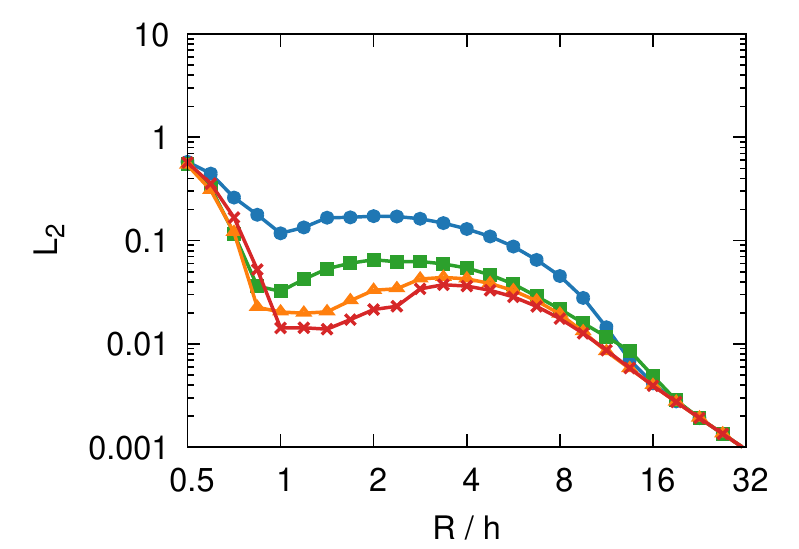}%
  \includegraphics{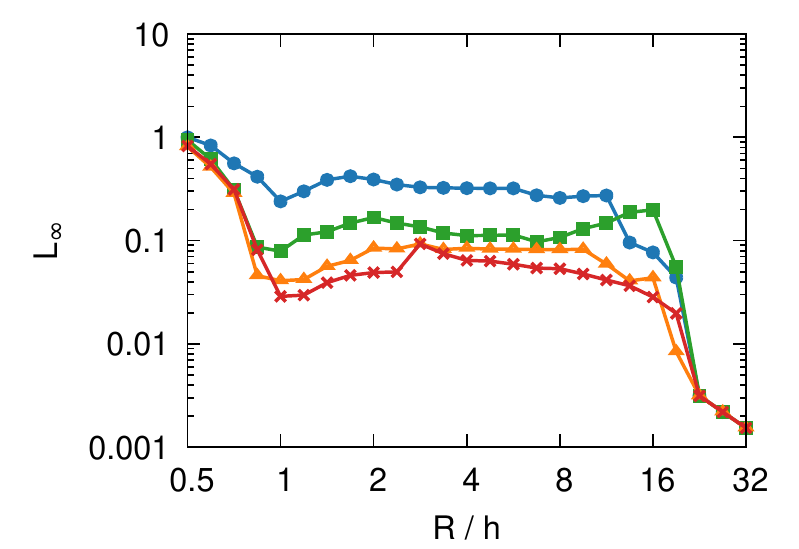}
  \caption{%
    Curvature error for a sphere in $L_2$ and $L_\infty$ norms
    depending on the resolution
    for various values of the particle string length
    $H_p=2$~\key{011}, 
    $3$~\key{032}, 
    $4$~\key{023} 
    and $5$~\key{044}.
  }
  \label{f:varyhp}
\end{figure*}

Estimation of the interface curvature in three dimensions depends
on three parameters:
the number of particles per string~$N$,
the number of cross sections~$N_s$
and the string length~$H_p$.
We set the parameters to $N=7$, $N_s=3$ and $H_p=4$.
Then, we vary each parameter independently and
examine their influence by estimating the curvature of a sphere
at various resolutions following the test case in Section~\ref{s:curv}.
The figures present the median error over 100 samples for the center
\revv{from a uniform distribution over the octant of the cell,
i.e. each coordinate is sampled from the range $[0,h/2]$}.
As seen from Figure~\ref{f:varynp},
the number of particles has a minor influence on the result. 
Nevertheless, we observe that $N=3$ 
provides a two times larger error for bubbles at resolutions about
one cell per radius. 
The influence of $N_s$ in Figure~\ref{f:varyns} 
is also small which is expected for a sphere.
However, more complex shapes such as those observed during the bubble
coalescence in Section~\ref{s:coal}, require at least $N_s=3$
as shown in Figure~\ref{f:coalns}.
Figure~\ref{f:varyhp} shows a stronger influence of the string length $H_p$.
Increasing the value from $H_p=2$ to $H_p=4$
reduces the error by a factor of ten.

\newcommand{\coalns}[2]{%
  \raisebox{2.5cm}{\hspace{1.4cm}$N_s=#2$\hspace{-2.7cm}}
  \includegraphics[width=0.25\textwidth,trim={6cm 12cm 6cm 12.2cm},clip]
  {coalns/#1}%
}

\begin{figure*}
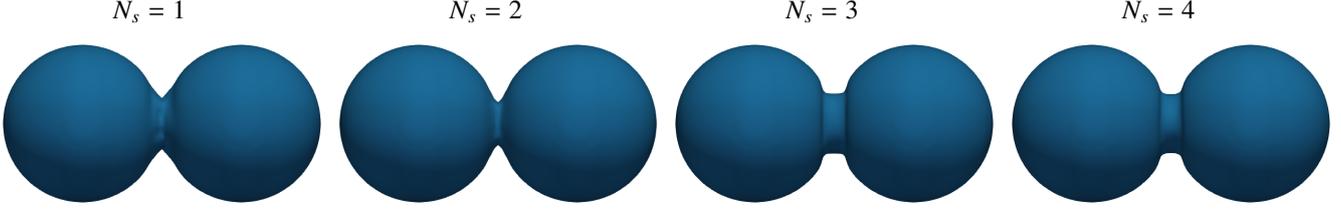

    \centering
    \coalns{ch128ns1/a.png}{1}
    \coalns{ch128ns2/a.png}{2}
    \coalns{ch128ns3/a.png}{3}
    \coalns{ch128ns4/a.png}{4}
    \caption{%
      Isosurfaces of the volume fraction with $R/h=19.2$ at $t/T=0.08$
      depending on the number of cross sections
      $N_s=1$, $2$, $3$ and $4$.
    }
    \label{f:coalns}
\end{figure*}

\subsection{\revvg{Computational cost}}

We compare the computational cost of our method
to that of the generalized height-function method
on the test case with a Taylor-Green vortex introduced in Section~\ref{s:tg}.
The formulation is modified to consider 890 spherical bubbles
at resolution $R/h=4$ on a mesh of $256^3$ cells.
The initial positions of bubbles are distributed uniformly in the domain.
The computations have been performed in parallel with 512 cores
on Piz Daint supercomputer,
where each compute node is equipped with a 12-core CPU Intel® Xeon® E5-2690 v3.
Table~\ref{t:timing} reports the runtime for both methods.
About 5\% of the cells contain interface fragments
and require the curvature estimation.
The time spent on the curvature estimation
with the present method amounts to 125~ms per step or 31.1\% of total time,
while GHF takes 33~ms per step or 10.2\% of total time.
Therefore, our method is about four times
more computationally expensive than GHF.

The cost of our method depends on the number of iterations
and the number of particles per cell,
which is controlled by parameters 
$N$, $N_s$ and $\varepsilon_p$.
Estimation of curvature for a given volume fraction field
consists of two parts: extraction of line segments and iterations.
Extraction of the line segments from neighboring cells
takes about 40\% of the time and scales linearly with $N_s$.
Iterations for equilibration of particles
take about 60\% of the time and scale linearly with $N_s\,N$.

\begin{table}
  \centering
  \begin{tabular}{p{0.62\columnwidth}|cc}
    \hline
    & present & GHF
    \\\hline
    total time, per step & 403 ms & 329 ms
    \\\hline
    curvature estimation time, per step & 125 ms & 33 ms
    \\\hline
    curvature estimation time, to total time & 31.1\% & 10.2\%
    \\\hline
    cells containing interface,
    to all cells & 5.17\% & 5.04\%
    \\\hline
    cells with curvature from height functions,
    to all cells & 0.69\% & 0.65\%
    \\\hline
    \hline
  \end{tabular}
  \caption{Runtime of the Taylor-Green vortex with 890 bubbles
  at resolution~$R/h=4$ on a mesh of $256^3$.
  The benchmark is executed on Piz Daint with 512 cores.
  The runtime of one time step is averaged over all steps,
  and the number of interfacial cells is measured at the final time $t=10$.}
  \label{t:timing}
\end{table}

\section{Test cases}
\label{s:test}

We examine the capabilities of the proposed method on two and three-dimensional 
benchmark problems: 
curvature of a sphere, a static droplet and a translating droplet.
The volume fraction is initialized by the exact volume 
cut by a sphere (circle)~\cite{strobl2016}.

\subsection{Curvature}
\label{s:curv}

The volume fraction field represents a single sphere (circle)
of radius $R$.
We vary the number of cells per radius $R/h$
and consider 100 samples for the center 
from a uniform distribution over the octant (quadrant)
of the cell, i.e. sampling each coordinate from the range $[0,h/2]$.
We compute the relative curvature error in $L_2$ and $L_\infty$ norms
\begin{align}
  L_2(\kappa) &= \Big( \frac{1}{|I|} \sum_{i\in I} 
  \Big(\frac{\kappa_i - \kappa_\text{ex}}{\kappa_\text{ex}}\Big)^2
  \Big) ^{1/2},
  \\ 
  L_\infty(\kappa) &= \max_{i \in I}
  \Big|\frac{\kappa_i - \kappa_\text{ex}}{\kappa_\text{ex}}\Big|,
\end{align}
where $I$ is the indices of cells containing the interface
(i.e. cells $i$ for which $0<\alpha_i<1$)
and $\kappa_\text{ex}$ is the exact curvature
(i.e. $\kappa_\text{ex}=2/R$ for sphere and 
$\kappa_\text{ex}=1/R$ for circle).

\begin{figure*}
  \centering
  \hspace*{\fill}%
  \includegraphics{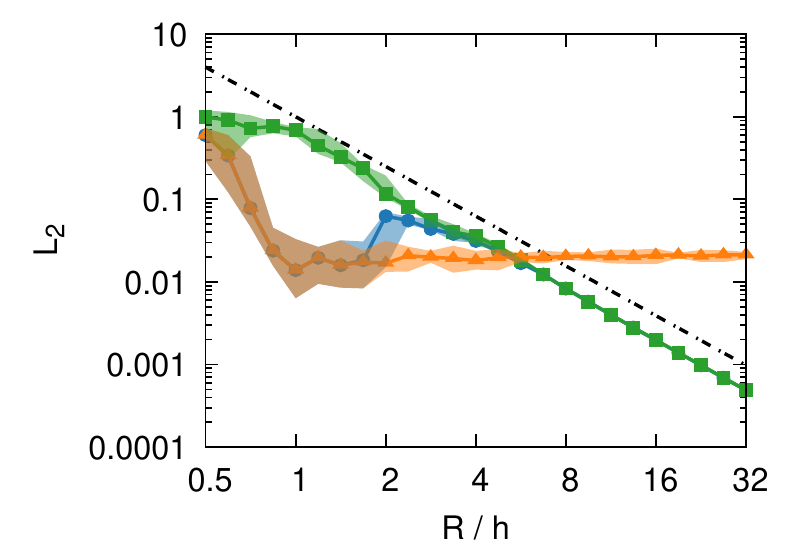}%
  \hspace*{\fill}%
  \includegraphics{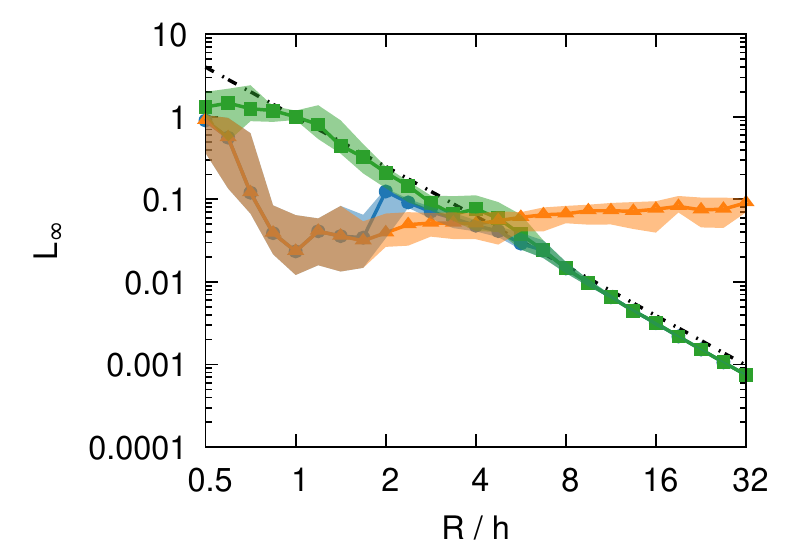}%
  \hspace*{\fill}%

  \hspace*{\fill}%
  \includegraphics{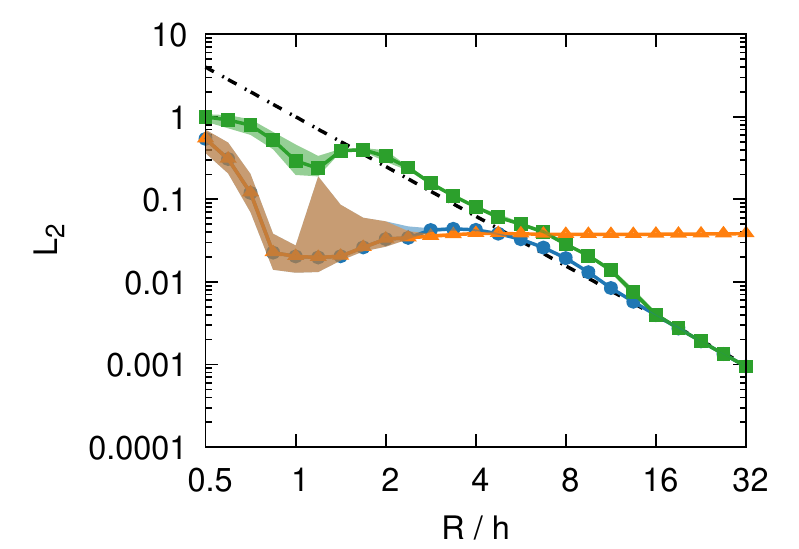}%
  \hspace*{\fill}%
  \includegraphics{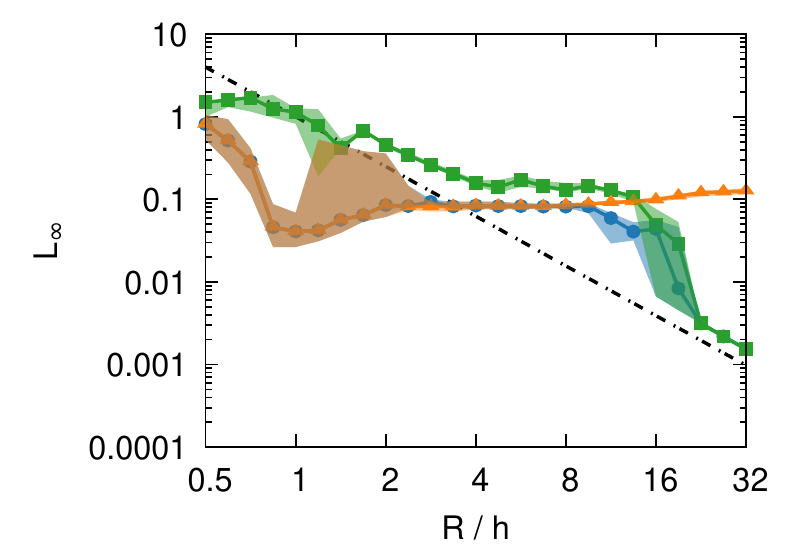}%
  \hspace*{\fill}%

  \caption{%
    Curvature error for a circle (top) and sphere (bottom)
    in $L_2$ and $L_\infty$ norms depending on the resolution:
    particles only~\key{023},
    combined particles and heights~\key{011}, GHF~\key{032}
    and second-order convergence~\key{200}.
    The lines show the median and the shades show the 
    10\% and 90\% percentiles for random positions of the center.
  }
  \label{f:curvoverlap}
\end{figure*}

Figure~\ref{f:curvoverlap} shows the error in comparison to GHF.
At low resolutions, our method is more accurate in terms of the $L_2$-error
up to eight cells per radius in 3D (and four cells in 2D)
and at resolutions below two cells per radius the error is 
a factor of ten smaller. 
\rev{The method of particles alone as described in Section~\ref{s:curv3d}
does not converge at high resolutions, and the 
maximum error saturates at about 10\%.
The combined method from Section~\ref{s:comb}
switches to height functions at high resolutions
and therefore converges with second order.}
Figures~\ref{f:randc}-\ref{f:randc3} show examples of
final configuration of particles at resolutions below
two cells per radius.

\newcommand{\randcplot}[1]{%
  \frame{\includegraphics[width=3cm]{randcplot/dim2_#1/vf_0000.pdf}}%
}

\newcommand{\randcplott}[1]{%
  \frame{\includegraphics[width=3cm]{randcplot/dim3_#1/a.png}}%
}

\begin{figure*}
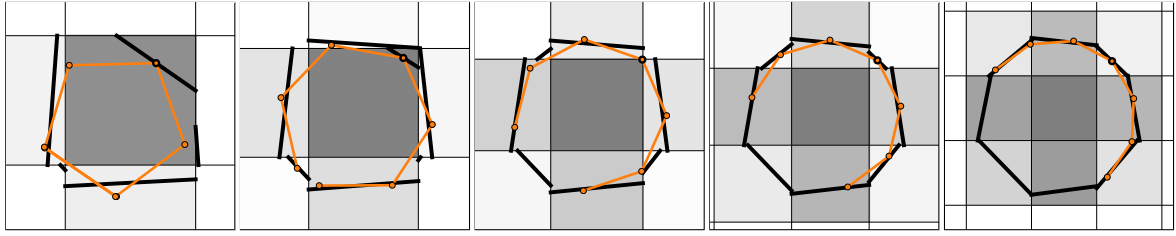

  \centering
  \randcplot{pltr001b000}
  \randcplot{pltr002b000}
  \randcplot{pltr003b000}
  \randcplot{pltr004b000}
  \randcplot{pltr005b000}
  \caption{Line segments of the interface (black)
   and positions of particles (orange) from one selected cell
   for a circle at resolutions 
   $R/h=0.59,\; 0.71,\; 0.84,\; 1.0\;\text{and}\;1.19$.
   The central particle is highlighted by a thicker edge.
  }
  \label{f:randc}
\end{figure*}

\begin{figure*}
  \centering
  \randcplott{pltr001b000}
  \randcplott{pltr002b000}
  \randcplott{pltr003b000}
  \randcplott{pltr004b000}
  \randcplott{pltr005b000}

  \caption{
    Polygons of the interface,
    cross sections (black) in case $N_s=2$
   and positions of particles (orange) from one selected cell
   for a sphere at resolutions
   $R/h=0.59,\; 0.71,\; 0.84,\; 1.0\;\text{and}\;1.19$.
  }
  \label{f:randc3}
\end{figure*}

\subsection{Static droplet}
\label{s:static}

We apply the model described in Section~\ref{s:flow}
for a spherical (circular) droplet in equilibrium.
The initial velocity is zero and the volume fraction field 
represents a single sphere (circle) of radius~$R$.
We assume that both components have the same density and viscosity
such that $\rho_1=\rho_2=\rho$ and $\mu_1=\mu_2=\mu$.
This leaves one dimensionless parameter, 
the Laplace number characterizing the ratio 
between the surface tension, inertial and viscous forces
\begin{equation}
\text{La}=\frac{2\sigma\rho R}{\mu^2},
\end{equation}
which we set to $\text{La}=1200$.
We solve the problem on a mesh of $128^3$ cells (or $128^2$ cells in 2D)
placing the droplet center in the corner and imposing the symmetry conditions
on the adjacent boundaries.  The other boundaries are free-slip walls.
We vary the number of cells per radius $R/h$
and advance the solution until time $T=\rho (2R)^2 / \mu$.

In the exact solution, the velocity remains zero
and the pressure experiences a jump at the interface given by 
the Laplace pressure $p_L = \sigma \kappa_\text{ex}$.
The numerical solutions develop spurious currents 
due to an imbalance between the pressure gradient and surface tension.
We compute the magnitude of the spurious velocity in the $x$-direction
\begin{equation}
  U_\text{max} = \max_{i\in C} |u_{x,i}|,
\end{equation}
the corresponding Weber number
\begin{equation}
  \label{e:wemax}
  \text{We}_\text{max} = \frac{2\rho R {U}_\text{max}^2}{\sigma} 
\end{equation}
and the pressure jump 
\begin{equation}
  \label{e:pjump}
  \Delta p = \max_{i\in C} p_i - \min_{i\in C} p_i,
\end{equation}
where $C$ is the indices of all cells.

\begin{figure*}
  \def\d{static/3d}
  \centering
  \hspace*{\fill}%
  \includegraphics{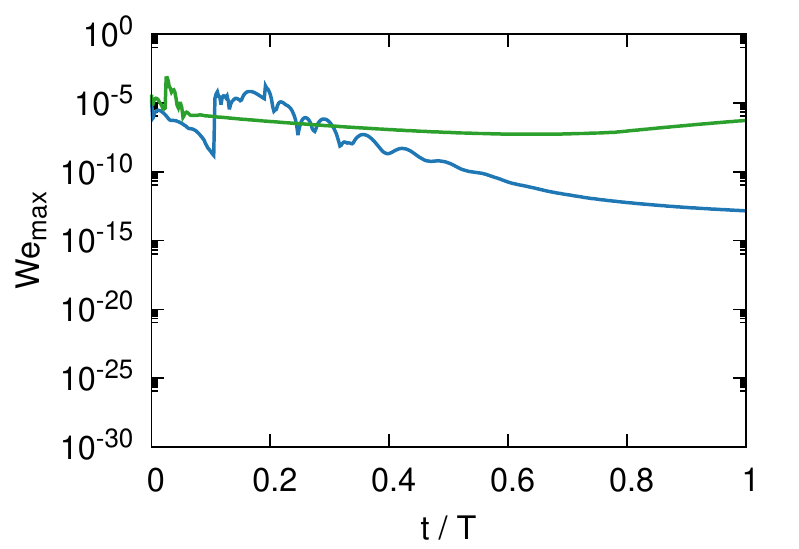}%
  \hspace*{\fill}%
  \includegraphics{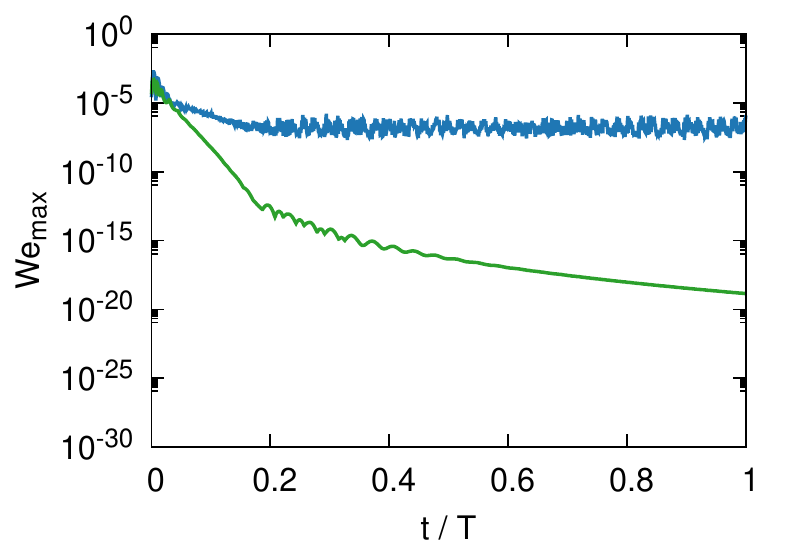}%
  \hspace*{\fill}%

  \caption{%
    Evolution of the velocity error 
    for a static droplet in 3D
    at resolutions $R/h=2$ (left) and $4$ (right): 
    present~\key{010} and GHF~\key{030}.
  }
  \label{f:st3traj}
\end{figure*}

Figures~\ref{f:st2}-\ref{f:st3} show the values of $\text{We}_\text{max}$
and the relative pressure jump $\Delta p / p_L$
at $t=T$ depending on the resolution.
The evolution of $\text{We}_\text{max}$ for 
two selected resolutions is shown in Figure~\ref{f:st3traj}.
We observe that the solutions from GHF 
convergence with time to a zero spurious flow \rev{in most cases.}
This demonstrates the existence 
of a volume fraction field for which the method of height functions
provides a uniform curvature field.
\rev{In other cases, such as $R/h=2$ and $26.9$ in 3D, 
the spurious flow does not converge to zero. 
Meanwhile, our method of particles more accurately predicts 
the pressure jump at low resolutions below 
$R/h=4$ cells in 3D (and 3.36 cells in 2D)
and provides lower magnitude of the spurious flow in cases 
such as $R/h=2$ in 3D shown in Figure~\ref{f:st3traj}.
}

\rev{We note, however, that the equilibration of a static droplet
is largely of theoretical interest
as the equilibrium shape is symmetric and aligned with the mesh directions.
Such conditions are incompatible with advection and therefore
rarely observed in practically relevant simulations.}
The spurious flow occurs in a more 
realistic scenario such as the translating droplet case 
discussed in Section~\ref{s:trans}.

\newcommand{\fig}{
  \centering
  \hspace*{\fill}%
  \includegraphics{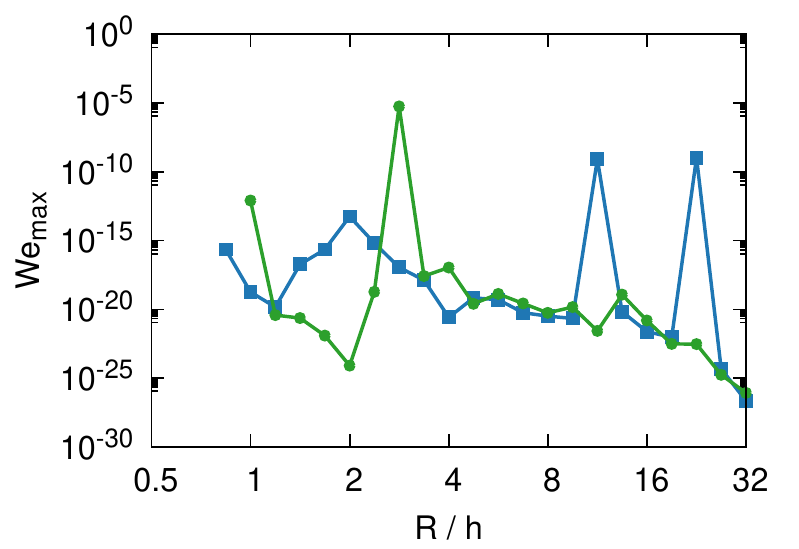}%
  \hspace*{\fill}%
  \includegraphics{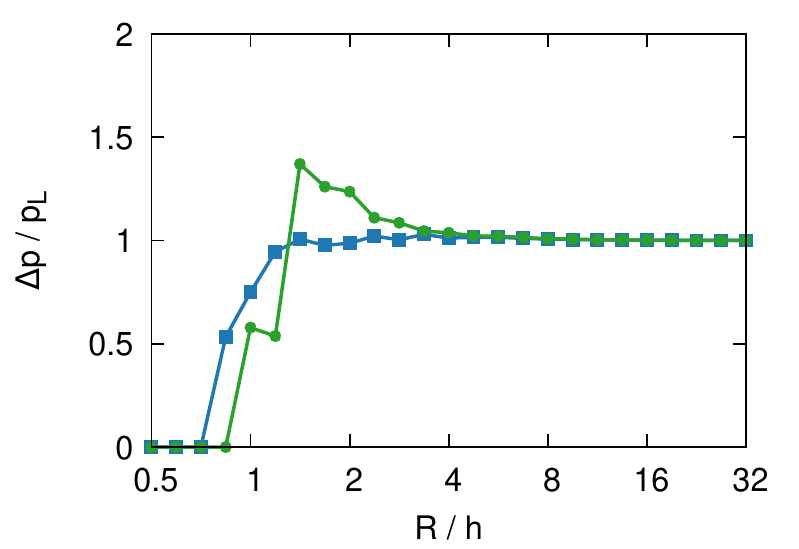}%
  \hspace*{\fill}%
}

\begin{figure*}
  \def\d{static/2d}
  \fig
  \caption{%
    Velocity error and the pressure jump 
    for a static droplet in 2D depending on the resolution:
    present~\key{012} and GHF~\key{031}.
  }
  \label{f:st2}
\end{figure*}

\begin{figure*}
  \def\d{static/3d}
  \fig
  \caption{%
    Velocity error and the pressure jump 
    for a static droplet in 3D depending on the resolution:
    present~\key{012} and GHF~\key{031}.
  }
  \label{f:st3}
\end{figure*}

\subsection{Translating droplet}
\label{s:trans}

We extend the previous case by adding a uniform 
initial velocity field
$\mb{u}(\mb{x},0)=\mb{U}=U\mb{l}/|\mb{l}|$ 
where $\mb{l}=(1,\;0.8,\; 0.6)$ in 3D and $\mb{l}=(1,\; 0.8)$ in 2D.
The additional parameter of the problem is the Weber number
\begin{equation}
  \text{We} = \frac{2\rho R U^2}{\sigma},
\end{equation}
which we set to $\text{We}=0.1$
while keeping $\text{La}=1200$.
We solve the problem in a periodic domain
on a mesh of $128^3$ cells (or $128^2$ cells in 2D).
We vary the number of cells per radius $R/h$
and advance the solution until time $T=2R/U$.

\begin{figure*}
  \def\d{univel/3d}
  \centering
  \hspace*{\fill}%
  \includegraphics{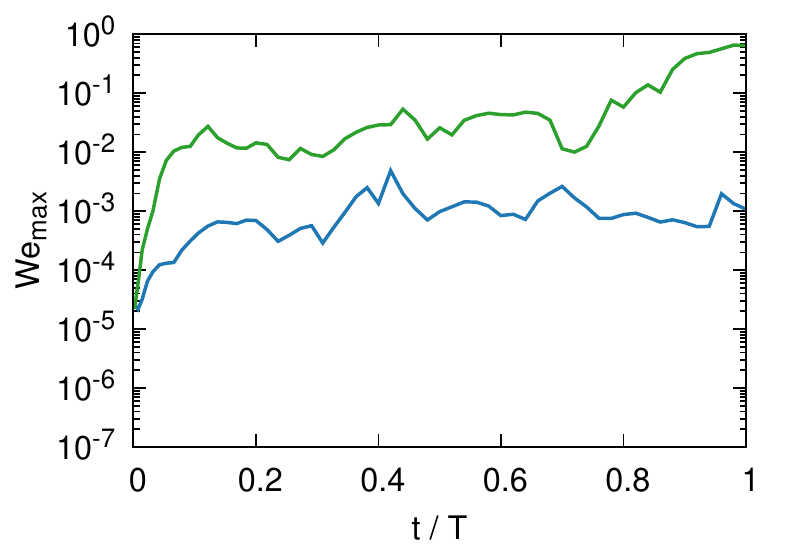}%
  \hspace*{\fill}%
  \includegraphics{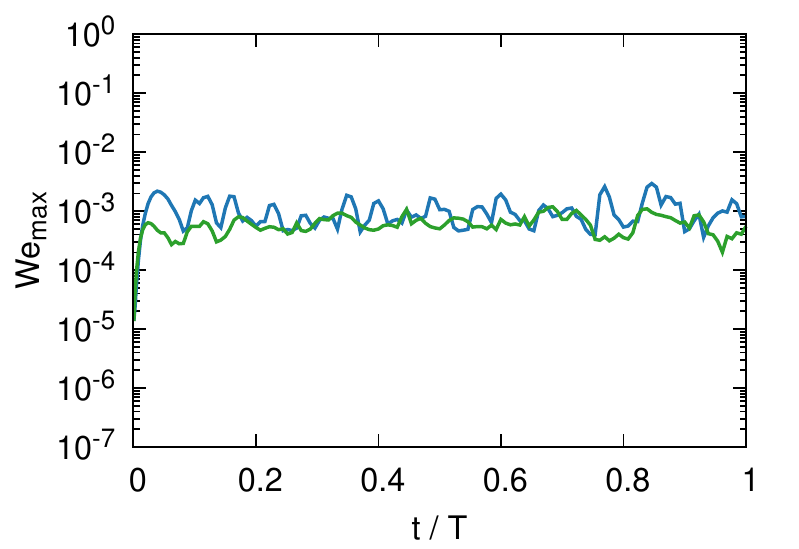}%
  \hspace*{\fill}%

  \caption{%
    Evolution of the velocity error
    for a translating droplet in 3D 
    at resolutions $R/h=2$ (left) and $4$ (right): 
    present~\key{010} and GHF~\key{030}.
  }
  \label{f:tr3traj}
\end{figure*}

The magnitude of the spurious flow is computed relative to
the initial velocity as the maximum over all cells
\begin{equation}
  U_\text{max} = \max_{i\in C} |u_{x,i}-U_x|
\end{equation}
and definitions of $\text{We}_\text{max}$ and $\Delta p$ 
follow~(\ref{e:wemax}) and (\ref{e:pjump}).
Figures~\ref{f:tr2}-\ref{f:tr3} show $\text{We}_\text{max}$
and the relative pressure jump $\Delta p / p_L$ depending on the resolution.
The quantities are averaged over $t\in[T/2,T]$
and the evolution of $\text{We}_\text{max}$ for 
two selected resolutions is shown in Figure~\ref{f:tr3traj}.
Our method provides lower magnitudes of the spurious flow than GHF
and more accurate values of the pressure jump
at resolutions below \rev{$R/h=2.82$ cells in 3D (and $R/h=2$ cells in 2D),
and at higher resolutions the error is comparable to GHF}.
\rev{We note that more accurate estimates of curvature in 
Section~\ref{s:curv} are observed at similar resolutions.}

\begin{figure*}
  \def\d{univel/2d}
  \fig
  \caption{%
    Velocity error and the pressure jump
    for a translating droplet in 2D depending on the resolution:
    present~\key{012} and GHF~\key{031}.
    Convergence with first~\key{100} and second~\key{200} order.
  }
  \label{f:tr2}
\end{figure*}

\begin{figure*}
  \def\d{univel/3d}
  \fig
  \caption{%
    Velocity error and the pressure jump
    for a translating droplet in 3D depending on the resolution:
    present~\key{012} and GHF~\key{031}.
    Convergence with first~\key{100} and second~\key{200} order.
  }
  \label{f:tr3}
\end{figure*}

\section{Applications}
\label{s:app}

\subsection{Taylor-Green vortex with bubble}
\label{s:tg}

The Taylor-Green vortex is a classical benchmark for 
the capabilities of flow solvers to
simulate single-phase turbulent flows~\cite{rees2011}.
Here we extend the formulation by adding a gaseous phase.
The problem is solved in a periodic domain $[0,2\pi]^3$
with the initial velocity
\begin{equation}
  \label{eq:tginit}
  \begin{split}
    u_x&=\sin x \;\cos y \;\cos z, \\
    u_y&=-\cos x \;\sin y \;\cos z, \\
    u_z&=0
  \end{split}
\end{equation}
and a single bubble of radius $R=0.1$ placed at $(2,2,2)$.
Parameters of the problem are 
the Reynolds number $\text{Re}=\rho_1/\mu_1$
and the Weber number $\text{We}=2\rho_1 R/\sigma$.
Here we choose $\text{Re}=800$ and $\text{We}=2$.
The density and viscosity ratios are set to
$\rho_2/\rho_1=0.01$ and $\mu_2/\mu_1=0.01$.

\begin{figure*}
  \centering
  \includegraphics{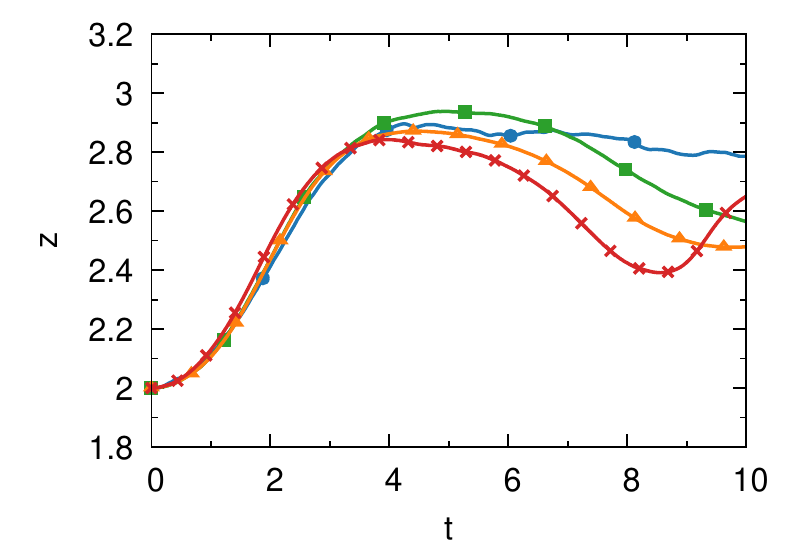}
  \includegraphics{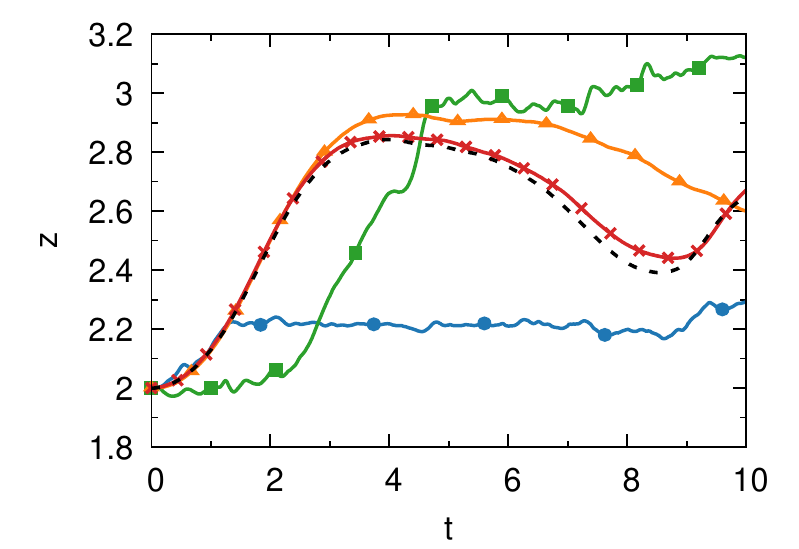}

  \caption{Trajectory of the bubble center of mass
  in the $z$-direction at various resolutions $R/h=$
    $1.53$~\key{011}, 
    $2.04$~\key{032}, 
    $3.06$~\key{023} 
    and $4.08$~\key{044}:
  present~(left) and GHF~(right).
  The dashed line~\key{100} on the right shows the trajectory 
  at $R/h=4.08$ from the present method.
  }
  \label{f:tgtraj}
\end{figure*}

\begin{figure*}
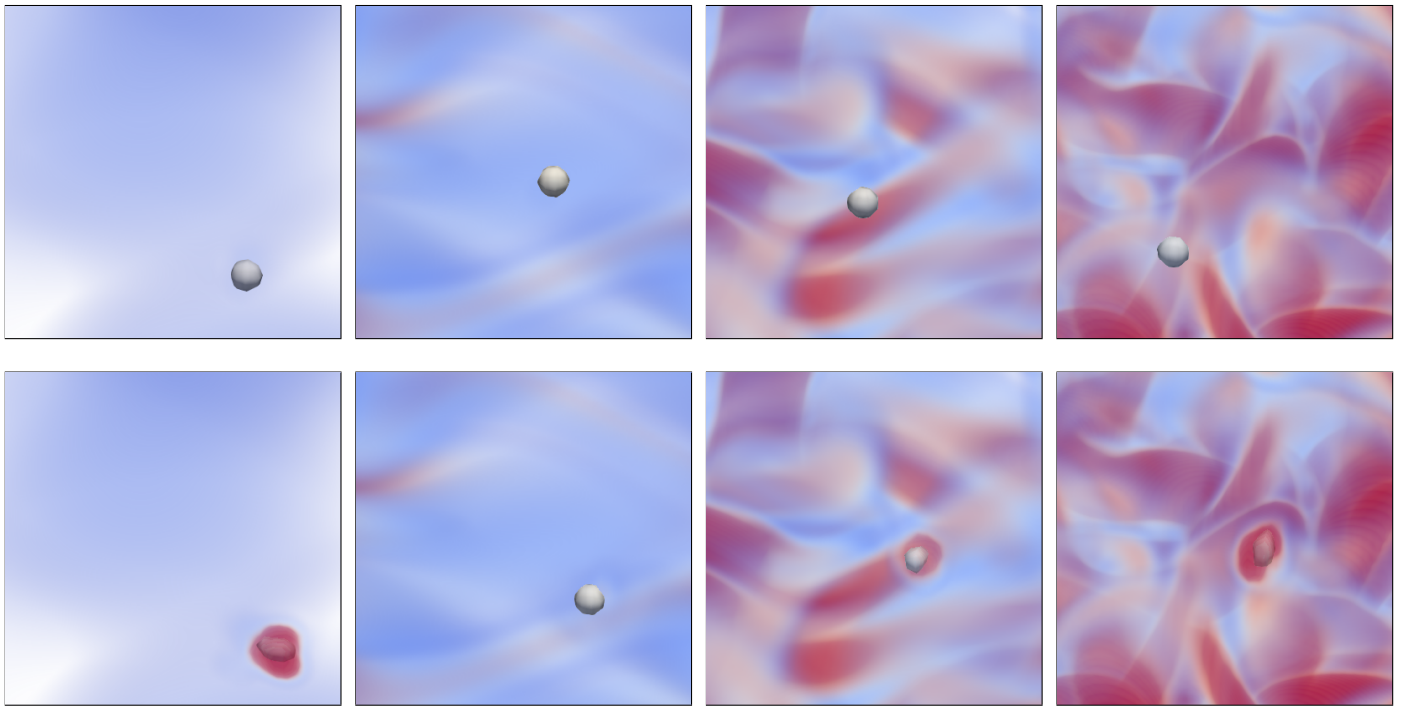

  \newcommand{\p}[1]{%
    \hfill
    \frame{\includegraphics[width=0.24\textwidth]{tg/view/#1}}%
    \hfill
  }
  \centering
  \p{bap/a_0002.png}%
  \p{bap/a_0007.png}%
  \p{bap/a_0012.png}%
  \p{bap/a_0017.png}%

  \bigskip

  \p{ba/a_0002.png}%
  \p{ba/a_0007.png}%
  \p{ba/a_0012.png}%
  \p{ba/a_0017.png}%

  \caption{Isosurfaces of the volume fraction and the magnitude of vorticity
  (increasing values from blue to red)
  with $R/h=2.04$ at $t=1.0,\;3.5,\;6.0\;\text{and}\;8.5$.
  \rev{At such resolution
  GHF (bottom) produces a solution with stronger spurious flow near the bubble
  than our method (top).}
  }
  \label{f:tgview}
\end{figure*}

Figure~\ref{f:tgtraj} shows the trajectory of the bubble 
at various resolutions in comparison to GHF.
Both methods converge to the same solution with the mesh refinement.
However, at lower resolutions $R/h\leq 3.06$
our method provides qualitatively more accurate trajectories,
while in GHF the trajectory is dominated by the spurious flow.
Snapshots of the vorticity magnitude and the bubble shape
are shown for both methods at  $R/h=3.06$ in Figure~\ref{f:tgview}

\subsection{\revvg{Taylor-Green vortex with elongated droplet}}

To evaluate our method on non-spherical shapes,
we consider the Taylor-Green vortex
with a droplet initially elongated in the $z$-direction.
We use the same initial conditions~(\ref{eq:tginit}) for the velocity field
and define the volume fraction to describe a single elliptical droplet
with~$R=0.1$
\begin{equation}
  \Big(\frac{x-2}{R}\Big)^2 + 
  \Big(\frac{y-2}{R}\Big)^2 + 
  \Big(\frac{z-2}{5R}\Big)^2 = 1.
\end{equation}
The density and viscosity ratios are set to
$\rho_2/\rho_1=10$ and $\mu_2/\mu_1=10$.
Other parameters of the problem are kept the same:
the Reynolds number $\text{Re}=\rho_1/\mu_1=800$
and the Weber number $\text{We}=2\rho_1 R/\sigma=2$.

\begin{figure*}
  \centering
  \includegraphics{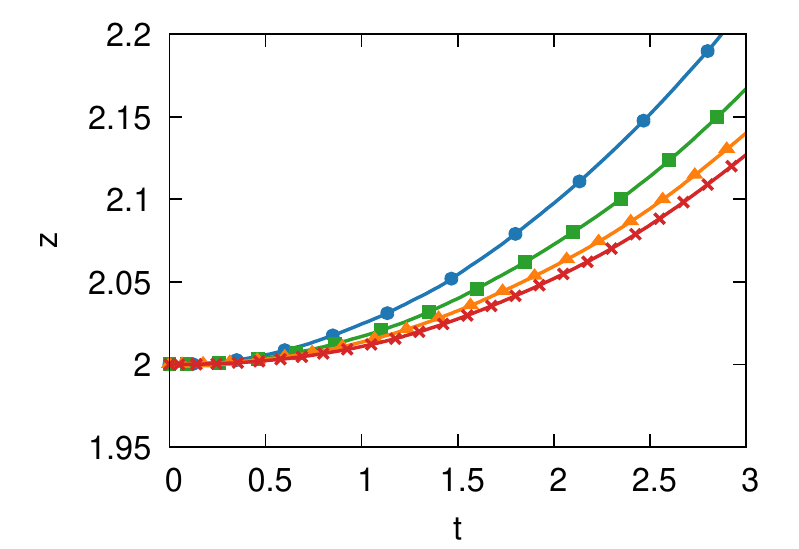}
  \includegraphics{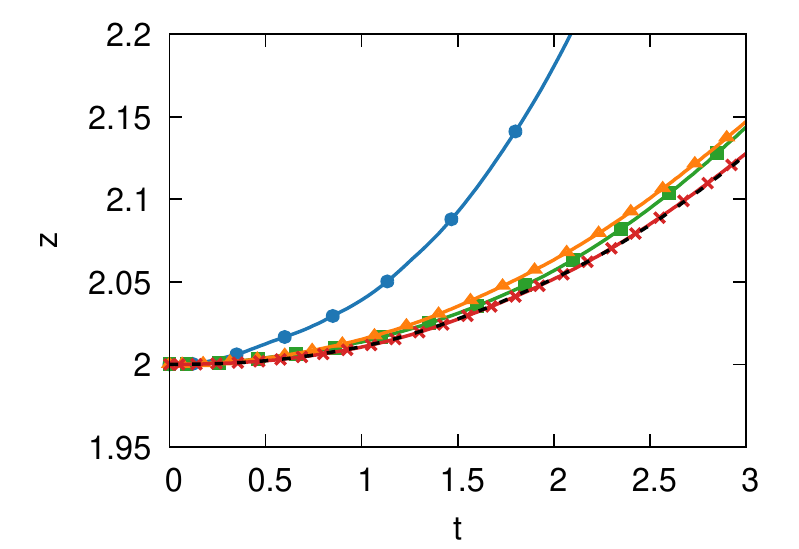}
  \includegraphics{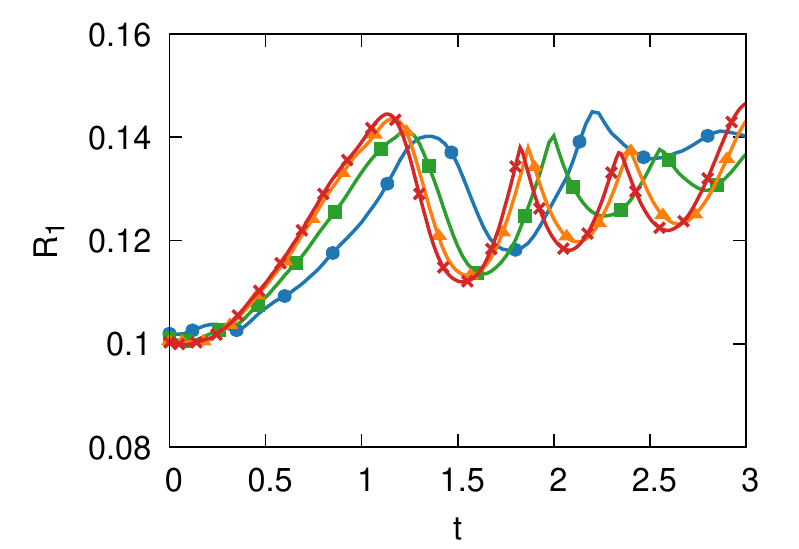}
  \includegraphics{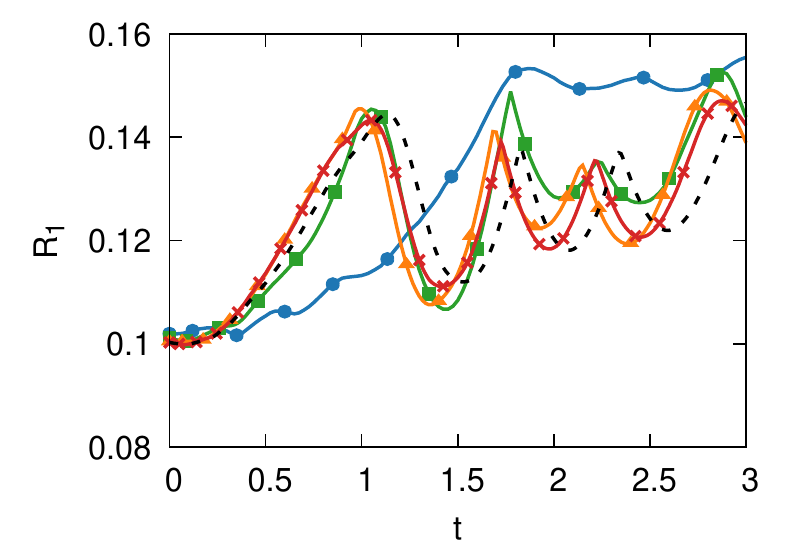}

  \caption{Trajectory of the droplet center of mass
  in the $z$-direction 
  and the smallest semi-axis of the ellipsoid of gyration
  at various resolutions $R/h=$
    $1.53$~\key{011}, 
    $2.04$~\key{032}, 
    $3.06$~\key{023} 
    and $4.08$~\key{044}:
  present~(left) and GHF~(right).
  The dashed line~\key{100} on the right shows the results
  at the finest resolution $R/h=4.08$ by the present method.
  }
  \label{f:tgtrajellip}
\end{figure*}

\begin{figure*}
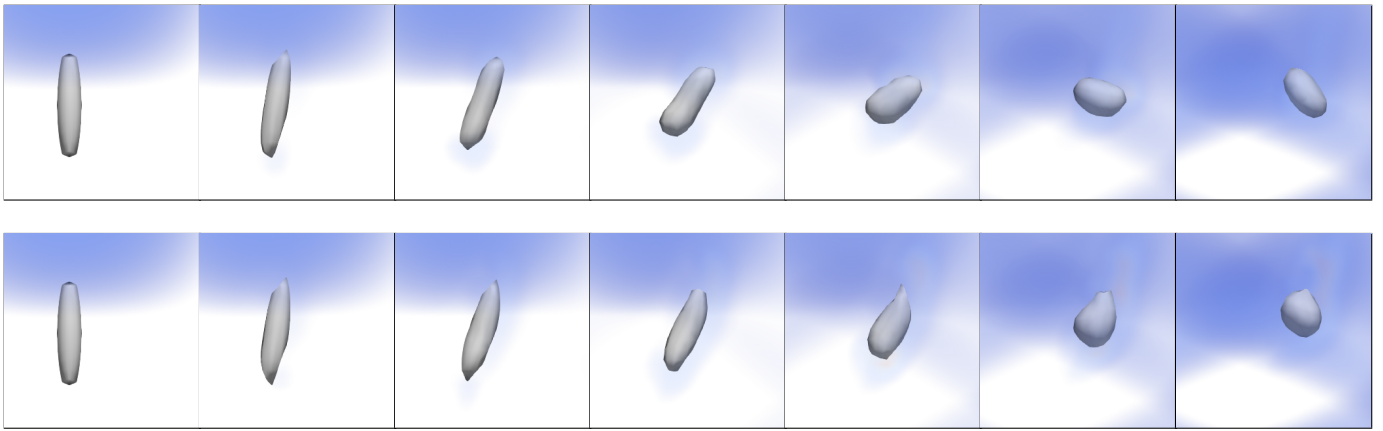

  \newcommand{\p}[1]{%
    \frame{\includegraphics[width=0.14\textwidth]{tgellipdrop/view/#1}}%
  }
  \centering
  \p{bap096/a_0000.png}%
  \p{bap096/a_0003.png}%
  \p{bap096/a_0006.png}%
  \p{bap096/a_0009.png}%
  \p{bap096/a_0012.png}%
  \p{bap096/a_0015.png}%
  \p{bap096/a_0018.png}%

  \bigskip

  \p{ba096/a_0000.png}%
  \p{ba096/a_0003.png}%
  \p{ba096/a_0006.png}%
  \p{ba096/a_0009.png}%
  \p{ba096/a_0012.png}%
  \p{ba096/a_0015.png}%
  \p{ba096/a_0018.png}%

  \caption{Isosurfaces of the volume fraction and the magnitude of vorticity
  (increasing values from blue to red)
  with resolution $R/h=1.53$ at $t=$0.0, 0.3, 0.6, 0.9, 1.2, 1.5 and 1.8:
  present method (top) and GHF (bottom).
  Unlike GHF, our method captures the oblate shape at $t=1.5$.
  }
  \label{f:tgviewellip}
\end{figure*}

\begin{figure*}
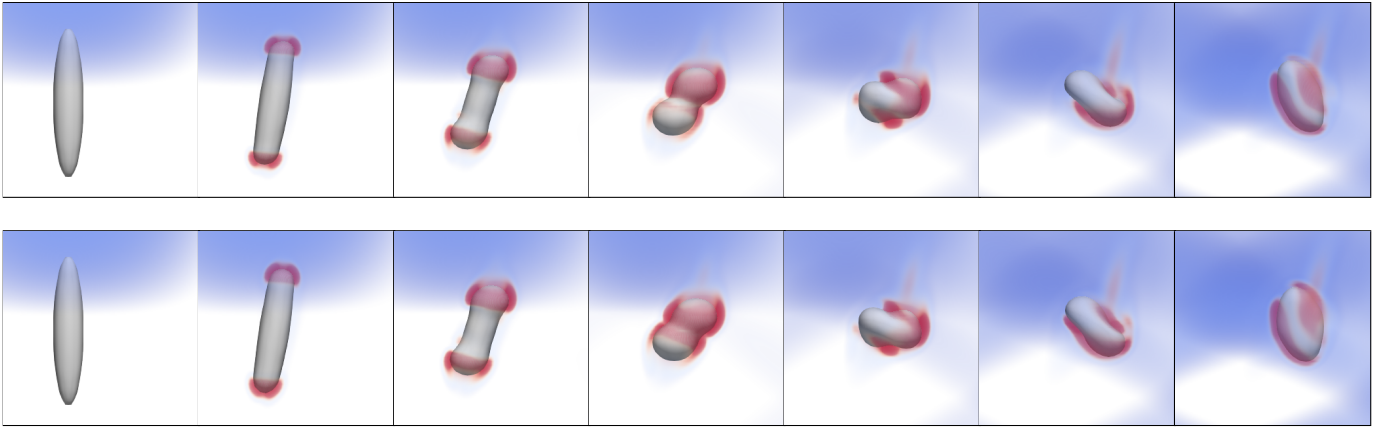

  \newcommand{\p}[1]{%
    \frame{\includegraphics[width=0.14\textwidth]{tgellipdrop/view/#1}}%
  }
  \centering
  \p{bap256/a_0000.png}%
  \p{bap256/a_0003.png}%
  \p{bap256/a_0006.png}%
  \p{bap256/a_0009.png}%
  \p{bap256/a_0012.png}%
  \p{bap256/a_0015.png}%
  \p{bap256/a_0018.png}%

  \bigskip

  \p{ba256/a_0000.png}%
  \p{ba256/a_0003.png}%
  \p{ba256/a_0006.png}%
  \p{ba256/a_0009.png}%
  \p{ba256/a_0012.png}%
  \p{ba256/a_0015.png}%
  \p{ba256/a_0018.png}%

  \caption{Isosurfaces of the volume fraction and the magnitude of vorticity
  (increasing values from blue to red)
  with resolution $R/h=4.08$ at $t=$0.0, 0.3, 0.6, 0.9, 1.2, 1.5 and 1.8:
  present method (top) and GHF (bottom).
  }
  \label{f:tgviewellip2}
\end{figure*}

The deformation of the droplet is characterized by the gyration tensor
\begin{equation}
  \frac{1}{V}\int (\mb{x}-\mb{x}_c)\otimes(\mb{x}-\mb{x}_c)\,\alpha\,dV,
\end{equation}
where $V=\int{\alpha\,dV}$ and $\mb{x}_c=\frac{1}{V}\int{\mb{x}\alpha\,dV}$.
For an elliptical droplet,
the principal components $\lambda_1<\lambda_2<\lambda_3$ 
of the gyration tensor
relate to semi-axes $R_1<R_2<R_3$ of the ellipsoid as 
$R_i=\sqrt{5\lambda_i},\;i=1,2,3$.
Figure~\ref{f:tgtrajellip} shows
the trajectory of the droplet center of mass in the $z$-direction
and the smallest semi-axis~$R_1$ of the ellipsoid of gyration
at various resolutions in comparison to GHF.
Snapshots of the vorticity magnitude and the bubble shapes
are shown in Figures~\ref{f:tgviewellip}-\ref{f:tgviewellip2}
for both methods at resolutions $R/h=1.53$ and $4.08$.
Both methods converge to the same solution with the mesh refinement.
The oscillation of the droplet is indicated by $R_1(t)$
which starts at $R_1=0.1$ according the initial conditions
and reaches the first maximum at $t=1.2$
when the droplet approaches a spherical shape shown in Figure~\ref{f:tgviewellip2}.
Due to inertia, the droplet further deforms to an oblate shape
corresponding to a minimum of $R_1$ at $t=1.5$.
At the lowest resolution $R/h=1.53$,
our method captures the formation of the oblate shape at $t=1.8$
which is delayed compared to the finest resolution.
However, with GHF the oblate shape is not captured.
At the next resolution $R/h=2.04$, 
GHF computes the trajectory more accurately than our method.

\subsection{Coalescence of bubbles}
\label{s:coal}

Coalescence of bubbles and drops is an actively studied phenomenon
commonly found in nature and industry~\cite{anthony2017,soto2018}.
Here we consider coalescence of two spherical bubbles.
The problem is solved in a periodic domain $[0,1]^3$
with zero initial velocity
and two tangent bubbles of radius $R=0.15$ placed along the $x$-axis.
The only parameter of the problem is the Ohnesorge number 
\begin{equation}
  \text{Oh} = \frac{\mu_1}{\sqrt{\rho_1R \sigma}},
\end{equation}
which we set to $\text{Oh}=0.007$.
The density and viscosity ratios are set to
$\rho_2/\rho_1=0.01$ and $\mu_2/\mu_1=0.01$.

We refer to~\cite{soto2018} for a detailed experimental
study of bubble coalescence.
The process starts with the formation of a neck connecting the bubbles
which then propagates along the bubble surface.
Figures~\ref{f:coalview}-\ref{f:coalview2} show
the isosurfaces of the volume fraction with different resolutions 
at $t/T=0.08$ and $0.18$,
where $T=\sqrt{\rho_1 R^3 / \sigma}$ is the capillary time.

We found that shapes from our simulations 
match well the experimental results (see Figures~\ref{f:coalview}, \ref{f:coalview2} and \ref{f:coalwall}) when a factor of 1.2 is applied to the values of time reported in~\cite{soto2018}. We note that the simulations using the boundary integral method~\cite{soto2018} capture the shapes reported in experiments without mentioning any such factor.
We performed additional simulations using  Gerris~\cite{popinet2009}
(see Figure~\ref{f:rnge}) and again matching shapes obtained by these simulations and those obtained experimentally required the adjustment by the factor of 1.2, in agreement with Basilisk and our own software. This discrepancy may be attributed to the use of an inviscid boundary integral method in~\cite{soto2018} and the viscous simulations employed in Gerris, Basilisk and our own software.
\revvg{
  Furthermore, we match the results of another
  experimental study~\cite{thoroddsen2005} without scaling of time
  as shown in Section~\ref{s:coalwall}
  and Figures~\ref{f:coalthor}-\ref{f:coalthorrn}.
}

\begin{figure*}
  \centering
  \includegraphics{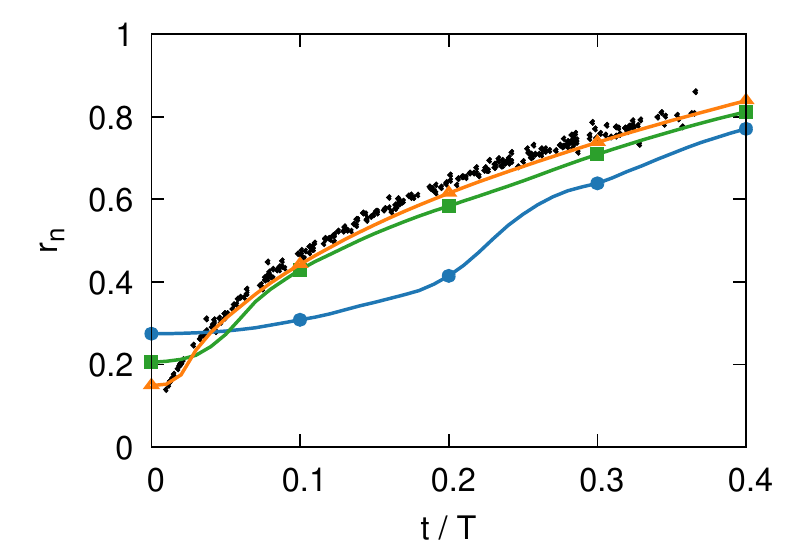}
  \includegraphics{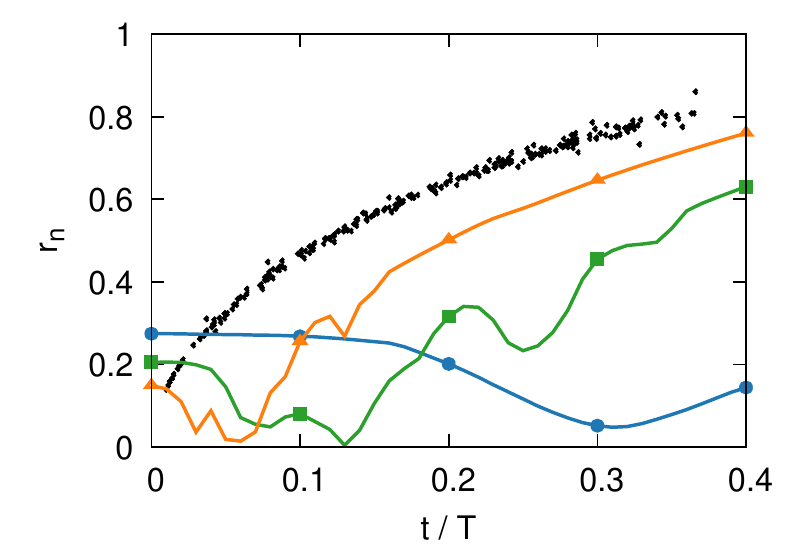}

  \caption{Evolution of the neck radius depending on the resolution
    $R/h=9.6$~\key{011}, 
    $19.2$~\key{032}, 
    $38.4$~\key{023}: present~(left) and GHF~(right)
    compared to 
    experiment~\cite{soto2018} (dots).
  }
  \label{f:coalrn}
\end{figure*}

\begin{figure}
  \centering
  \includegraphics{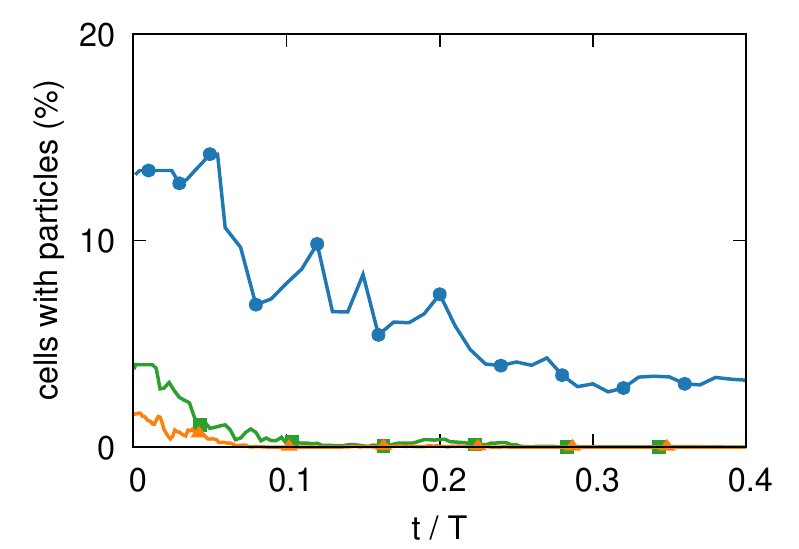}

  \caption{Percentage of interfacial cells where the curvature 
  is computed with particles for 
  $R/h=9.6$~\key{011}, $19.2$~\key{032}, $38.4$~\key{023}.
  }
  \label{f:coalhf}
\end{figure}

The evolution of the neck radius is presented 
in Figure~\ref{f:coalrn}
in comparison to the experimental data reported in~\cite{soto2018}.
We observe that the present method agrees well with experimental data 
while this is not the case for simulations using the generalized height function method, in particular at later times (see Figure \ref{f:coalview2}).
The generalized height-function method introduces spurious 
disturbances of the interface near the coalescence neck.
Furthermore, its solution does not converge with mesh refinement.
The reason for this is that
the initial shape has effectively infinite curvature at the coalescence neck
as refining the mesh increases the curvature resolved on the mesh.
\revvg{
As seen from Figure~\ref{f:coalhf},
increasing the resolution reduces the percentage
of cells where height functions are not defined
and the curvature is estimated using particles.
Nevertheless, more accurate estimation of curvature
in such cells benefits the overall accuracy.
}

\newcommand{\coal}[1]{
  \includegraphics[width=0.25\textwidth,trim={6cm 12cm 6cm 12cm},clip]
  {coal/#1}%
}

\newcommand{\coalexp}[1]{
  \includegraphics[width=0.25\textwidth,trim={1.6cm 2cm 1.6cm 6cm},clip]
  {coal/exp/#1}%
}

\begin{figure*}
    \centering
    \coal{ch064/a_0008.png}%
    \coal{ba064/a_0008.png}

    \coal{ch128/a_0008.png}%
    \coal{ba128/a_0008.png}

    \coal{ch256/a_0008.png}%
    \coal{ba256/a_0008.png}

    \coalexp{edge/edge_ch_0.png}%
    \coalexp{edge/edge_ba_0.png}

    \caption{Isosurfaces of volume fraction at $t/T=0.08$
    depending on resolution
    $R/h=9.6$ (top), $19.2$ (middle) and $38.4$ (bottom):
    present (left) and GHF (right).
    Experimental image~\cite{soto2018} overlapped with contours 
    from the finest mesh of each method.
    The time value for the experimental image is multiplied by 1.2.
    }
    \label{f:coalview}
\end{figure*}

\begin{figure*}
    \centering
    \coal{ch064/a_0018.png}%
    \coal{ba064/a_0018.png}

    \coal{ch128/a_0018.png}%
    \coal{ba128/a_0018.png}

    \coal{ch256/a_0018.png}%
    \coal{ba256/a_0018.png}

    \coalexp{edge/edge_ch_1.png}%
    \coalexp{edge/edge_ba_1.png}

    \caption{Isosurfaces of volume fraction at $t/T=0.18$
    depending on resolution
    $R/h=9.6$ (top), $19.2$ (middle) and $38.4$ (bottom):
    present (left) and GHF (right).
    Experimental image~\cite{soto2018} overlapped with contours 
    from the finest mesh of each method.
    The time value for the experimental image is multiplied by 1.2.
    }
    \label{f:coalview2}
\end{figure*}

\subsection{\revv{Breakup of air sheet in shear flow}}
\label{s:shear}

\begin{figure}
    \centering
    \begin{overpic}[unit=1mm,scale=0.172]{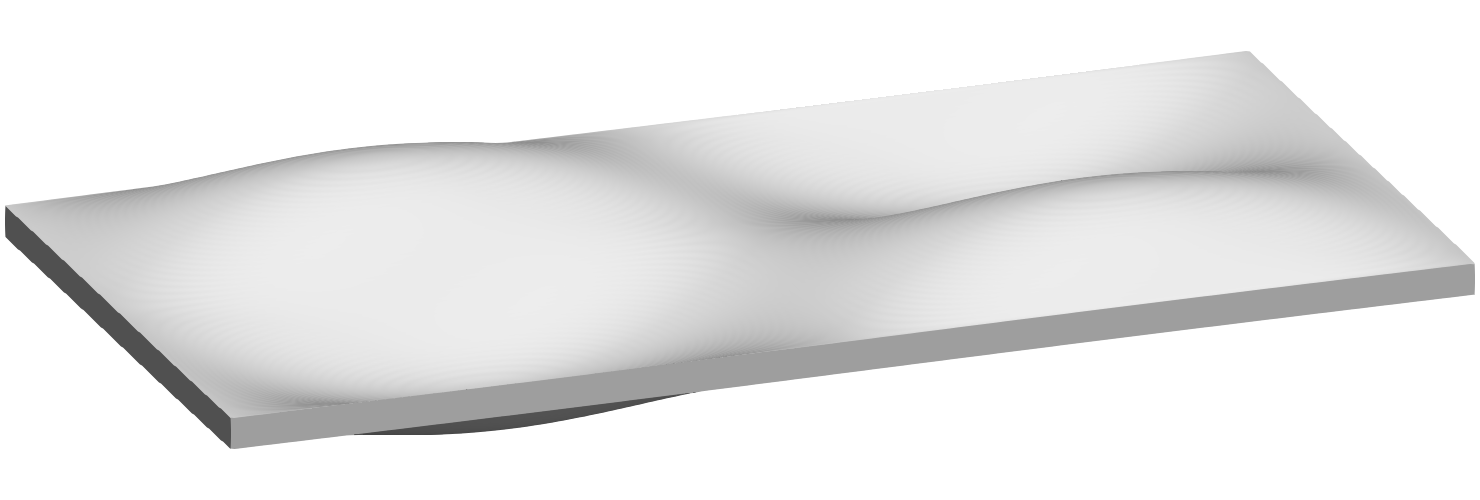}
      \put(0,22){\includegraphics[width=2cm]{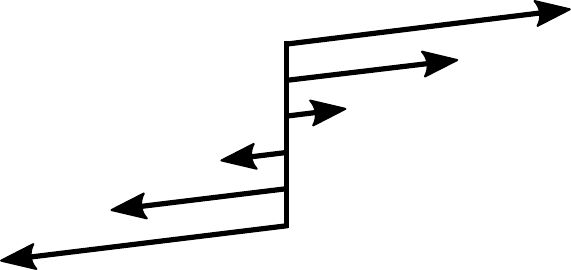}}
    \end{overpic}
    \caption{Initial volume fraction representing an air sheet.
             Arrows indicate the direction of the shear.}
    \label{f:shearviewinit}
\end{figure}

The following test case demonstrates the deformation
and breakup of an air sheet in shear flow.
The domain $[0,2]\times[0,1]\times[0,1]$ 
is periodic in the $x$- and $z-$directions
and no-slip conditions are imposed on the boundaries in the $y$-direction.
The initial velocity profile is simple shear
$u_x=(y - 0.5)U/D$ and $u_y=u_z=0$,
where $D=0.05$ is the thickness of the sheet
and $U=0.2$ is the velocity difference.
The initial volume fraction
\begin{equation}
  \alpha =
  \begin{cases}
    1, &\big|\,0.1 \sin(\pi x) \sin(2\pi z) + 0.5 - y \big| < \frac{D}{2} \\
    0, &\text{otherwise}
  \end{cases}
\end{equation}
is shown in Figure~\ref{f:shearviewinit}.
Parameters of the problem are 
the Reynolds number $\text{Re}=\rho_1 U D /\mu_1$
and the Weber number $\text{We}=\rho_1 U^2 D/\sigma$.
Here we choose $\text{Re}=80$ and $\text{We}=0.64$.
The density and viscosity ratios are set to
$\rho_2/\rho_1=0.01$ and $\mu_2/\mu_1=0.01$.

\begin{figure*}
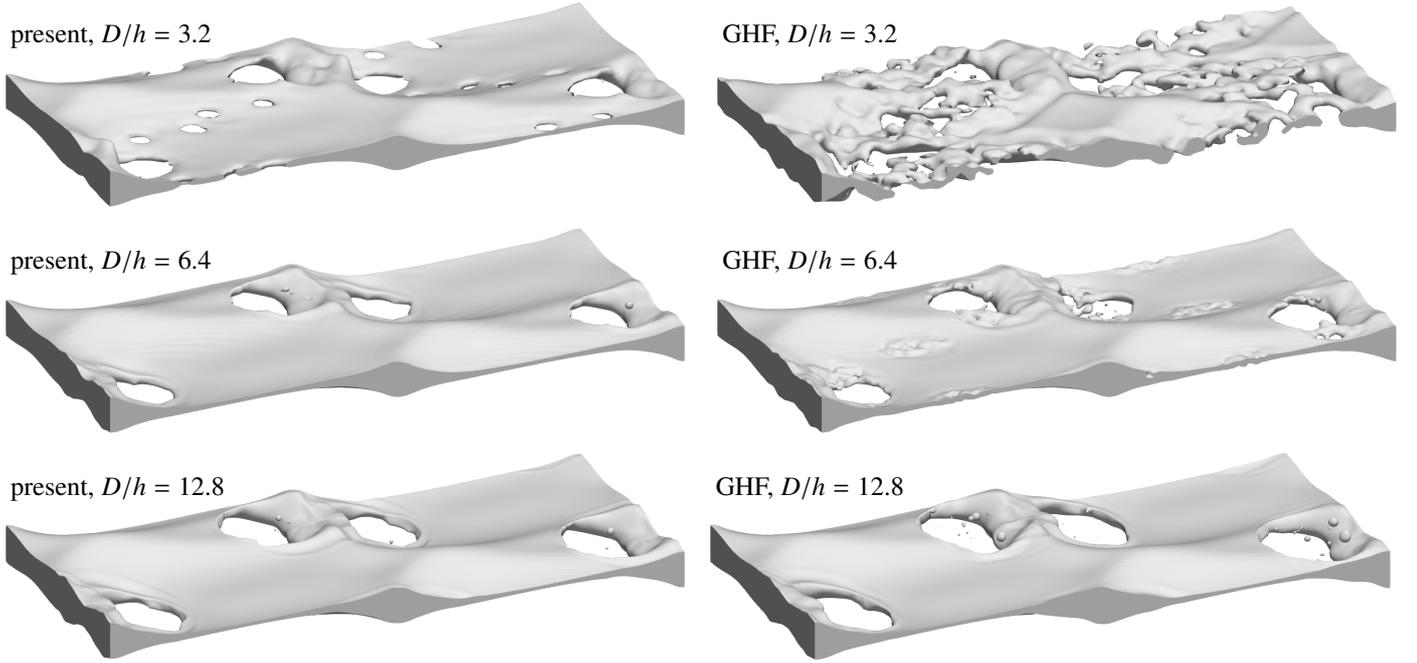

    \newcommand{\p}[2]{%
      \begin{overpic}[unit=1mm,scale=0.172]{shear/view/#1}
      \put(1,26){#2}
      \end{overpic}
    }
    \centering
    \p{bap128/a_0017.png}{present, $D/h=3.2$}%
    \hfill
    \p{ba128/a_0017.png}{GHF, $D/h=3.2$}
    \p{bap256/a_0017.png}{present, $D/h=6.4$}%
    \hfill
    \p{ba256/a_0017.png}{GHF, $D/h=6.4$}
    \p{bap512/a_0017.png}{present, $D/h=12.8$}%
    \hfill
    \p{ba512/a_0017.png}{GHF, $D/h=12.8$}

    \caption{Shapes of the air sheet at $t=1.7$
    computed with present method (left) and  GHF (right)
    at resolutions $D/h=$3.2 (top), 6.4 (middle) and 12.8 (bottom).}
    \label{f:shearview}
\end{figure*}

Initial deformations of the air sheet
develop further in the shear flow.
This leads to tearing of the interface starting in four distinct locations.
Figure~\ref{f:shearview}
shows the shapes computed with both methods of curvature estimation
at various resolutions.
Both methods converge to the same solution with mesh refinement.
At lower resolutions, GHF shows more tearing of the interface,
while with the present method the interface remains stable.

\section{Conclusion}
\label{s:concl}

We have presented a new method for estimating the curvature
of the interface by fitting circular arcs to its 
piecewise linear reconstruction.
The circular arcs are represented as strings of  particles
which evolve under constraints and forces attracting them to the interface.

The application of the method on a number of benchmark problems 
shows a significant improvement 
in the accuracy of computing the curvature of interfaces at low resolutions 
over the generalized height-function method~\cite{popinet2009}
implemented in Basilisk~\cite{basilisk}.
The present method is more accurate at resolutions up to \rev{four cells} 
per curvature radius
and even with one cell per radius 
provides the relative curvature error below 10\%.
We also demonstrate the capabilities of this hybrid method on a number of applications 
including multiphase vortical flows and bubble coalescence.
Further applications include
the bubble dynamics in electrochemical cells~\cite{hashemi2019}
and a plunging jet with air entrainment~\cite{pasc}.

The present  technique restricts the particles to circular arcs
and computes the attraction force from the nearest point
on the interface.
However, the method allows for modifications of the constraints and forces.
For instance, the force can be computed directly from the volume fraction
using the area cut by the string of particles.
Forces defined from the intersection with the reconstructed volume
can lead to an approach similar to the mesh-decoupled height 
functions~\cite{owkes2015}.
Finally, the property of recovering the exact curvature
mentioned in \rev{Section~\ref{s:circ}}
can contribute to the existence of volume fraction fields
providing a uniform curvature field and, therefore, 
exact equilibration of a static droplet.
Such modifications constitute the subject of future work.

\section{Acknowledgements}

This research is funded by grant no. \verb`CRSII5_173860` of the Swiss National Science Foundation.
The authors acknowledge the use of computing resources 
from CSCS (projects s754 and s931).
We thank Professor Georges-Henri Cottet (Grenoble, France) for several helpful discussions regarding this work.


\appendix
\section*{Appendix A}
\renewcommand{\thesubsection}{\Alph{section}\arabic{subsection}}
\renewcommand{\thefigure}{\arabic{figure}}
\setcounter{section}{1}

\subsection{Attraction to circular arcs}
\label{s:circ}

\begin{figure}
  \centering
  \includegraphics[scale=0.75]{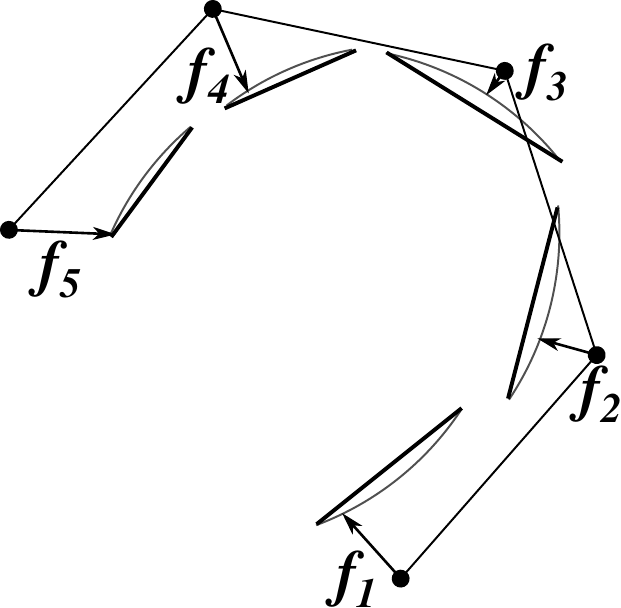}

  \caption{%
    Line segments represent the interface and
    circular arcs of curvature~$\kappa$ pass through their endpoints.
    The force acts on each particle towards the circular arc
    from the nearest line segment.
  }
  \label{f:sk_forcecirc}
\end{figure}

The attraction force~(\ref{e:force})
can be modified to include a dependency
on the current estimate of curvature.
One such modification consists in replacing the line segments
with circular arcs.
This formulation recovers the exact curvature
if the endpoints of all line segments belong to a circle.
To define the force at position~$\mb{x}$, we find the nearest
point~$\mb{y}\in L$ on the interface and the corresponding line
segment~$[\mb{a}_l,\mb{b}_l]$.  Then we find a factor~$\delta$ such that
\begin{equation}
  \label{e:forcedelta}
\mb{x}_L(\mb{x},\kappa) = 
  \mb{y} + \delta \mb{n}_l
\end{equation}
belongs to a circular arc of curvature~$\kappa$
through the endpoints $\mb{a}_l$ and $\mb{b}_l$,
where~$\kappa$ is a known estimation of curvature
and $\mb{n}_l$ is the unit normal of $[\mb{a}_l,\mb{b}_l]$.
Such~$\delta$ is given by
\begin{equation}
  \delta 
  = 
    \sqrt{\tfrac{1}{\kappa^2} - d^2} -
    \sqrt{\tfrac{1}{\kappa^2} - w^2}
  = 
    \frac{\kappa(w^2-d^2)}{ \sqrt{1 - \kappa^2d^2} + \sqrt{1-\kappa^2w^2}},
\end{equation}
where $d=|\mb{y}-\mb{c}_l|$, $w=|\mb{a}_l-\mb{c}_l|$ and 
$\mb{c}_l=(\mb{a}_l+\mb{b}_l)/2$.
Finally, the force is defined as
\begin{gather}
  \label{e:forcecirc}
  \mb{f}(\mb{x},\kappa) =
  \eta \; \big(\mb{x}_L(\mb{x},\kappa) - \mb{x} \big),
\end{gather}
where $\eta\in [0,\,1]$ is a relaxation parameter.
Figure~\ref{f:sk_forcecirc} illustrates the computation of forces
after replacing the line segments with circular arcs.

\subsection{Coalescence of bubbles with SIMPLE-based solver}
\label{s:coalwall}
The method for curvature estimation is also implemented 
as part of our in-house multiphase flow solver
\mbox{Aphros}.
We use a finite volume discretization based on
the SIMPLE method for the pressure coupling~\cite{patankar1983,ferziger2012}
and the second-order scheme QUICK~\cite{leonard1979} for convective fluxes.
The advection equation is solved using the volume-of-fluid method PLIC
with piecewise linear reconstruction~\cite{aulisa2007},
where the normals are computed using the mixed Youngs-centered scheme
which is a combination of Youngs' scheme and the height functions.
The cell-centered interface curvature 
$\kappa$ is computed from the volume fraction
using the proposed method.
Our approximation of the surface tension force is well-balanced 
\cite{francois2006}
(i.e.\ the surface tension force is balanced by the pressure gradient
if the curvature is uniform)
and requires face-centered values of the curvature
which are taken from the neighboring cell
with minimal~$|\alpha-\tfrac{1}{2}|$.

The algorithm is implemented on top of
Cubism~\cite{wermelinger2018,cubism}, an open-source C++ framework for
distributed parallel solvers on structured grids.  
To solve the linear systems, we use
the GMRES method~\cite{saad1986} for the momentum equation and 
the preconditioned conjugate gradient
method~\cite{ashby1996} for the pressure correction
implemented in the Hypre library~\cite{hypre,falgout2002}.

One functionality of our code
is the ability to describe static contact angles of $0$
and $180$ degrees.
This feature is implemented using ghost cells.
Assuming that $\alpha=1$ is the volume fraction inside the bubble or droplet,
we fill two layers of ghost cells adjacent to boundaries 
with values $\alpha=0$ and $\alpha=1$ for the contact angles
of $180$ and $0$ degrees respectively.
Based on such volume fraction field,
we estimate the normals and curvature in all domain cells with
the usual algorithm.
This allows us to describe the evolution of the two bubbles 
after coalescence presented in Section~\ref{s:coal}
at later stages as well and include the detachment from the solid wall
and oscillations.
The problem formulation is closer to the experimental 
conditions than in Section~\ref{s:coal}.
The bubbles are initially placed near a solid wall
with the imposed contact angle of 180 degrees
and the gravitational acceleration is considered
with the E{\"o}tv{\"o}s number of $\text{Eo}=\rho g R^2/\sigma=0.061$.
Results of the simulation compared to the experimental images
are shown in Figure~\ref{f:coalwall}.

\begin{figure}
  \centering
  \includegraphics{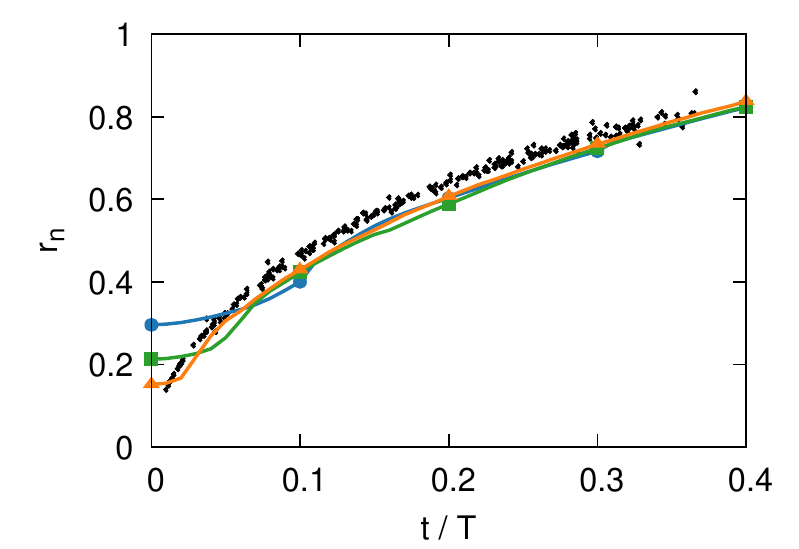}

  \caption{Evolution of the coalescence neck radius
  relative to the bubble radius depending on the resolution
    $R/h=9.6$~\key{011}, 
    $19.2$~\key{032}
    and $38.4$~\key{023} 
    produced by Aphros
    compared to experiment~\cite{soto2018} (dots).
  }
  \label{f:coalrnch}
\end{figure}

Furthermore, the evolution of the coalescence neck radius
on the coarsest resolution
is more accurate than in Basilisk as shown in 
Figures~\ref{f:coalrn} and~\ref{f:coalrnch}.
This is attributed to differences in the algorithm for computation of 
interface normals and the way the estimates of curvature
are transferred from cells to faces (average over neighbors in Basilisk,
and the cell with the minimal $|\alpha-\tfrac{1}{2}|$ in Aphros).

\begin{figure}
  \centering
  \def\coalvspace{\vspace{0.5mm}}
  \newcommand{\p}[1]{%
    \hfill%
    {\includegraphics[width=0.245\columnwidth]{coalthor/view/a_00#1.png}}%
    \hfill%
  }


  \p{00}%
  \p{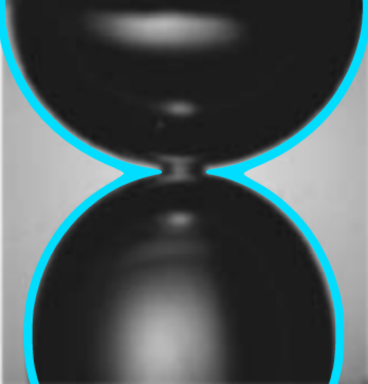}%
  \p{02}%
  \p{03}

  \coalvspace

  \p{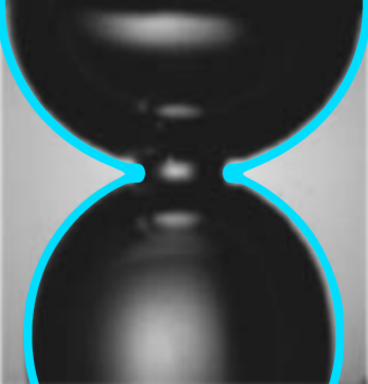}%
  \p{07}%
  \p{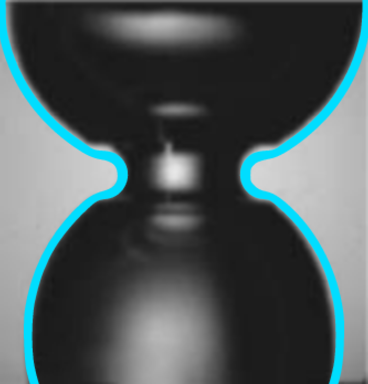}%
  \p{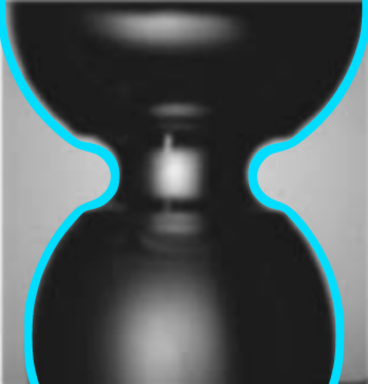}

  \coalvspace

  \p{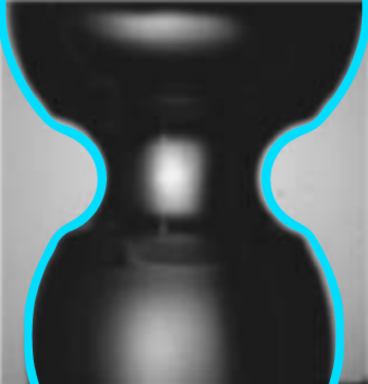}%
  \p{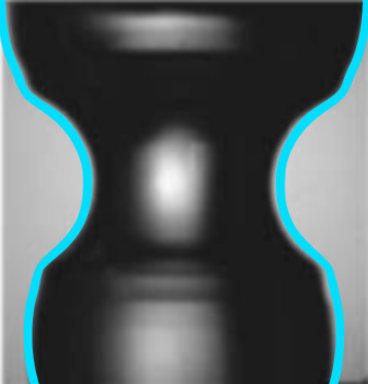}%
  \p{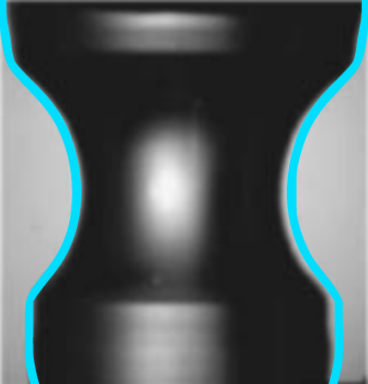}%
  \p{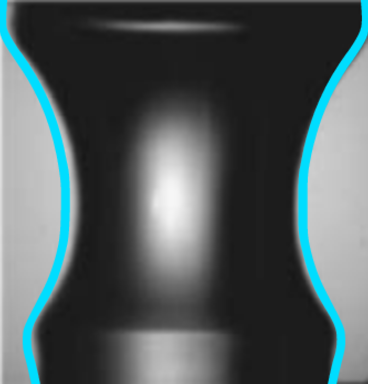}

  \caption{Snapshots from experiment~\cite{thoroddsen2005}
  overlaid by the projections of the shapes from the simulation
  by Aphros with resolution $R/h=76.8$ (blue lines)
  at times $t/T=$0, 0.0061, 0.012, 0.018, 0.031, 0.043, 0.061,
0.08, 0.12, 0.17, 0.21 and 0.27.
  Experimental images reprinted from~\cite{thoroddsen2005},
  with the permission from AIP Publishing.
  }
  \label{f:coalthor}
\end{figure}

\begin{figure}
  \centering
  \includegraphics{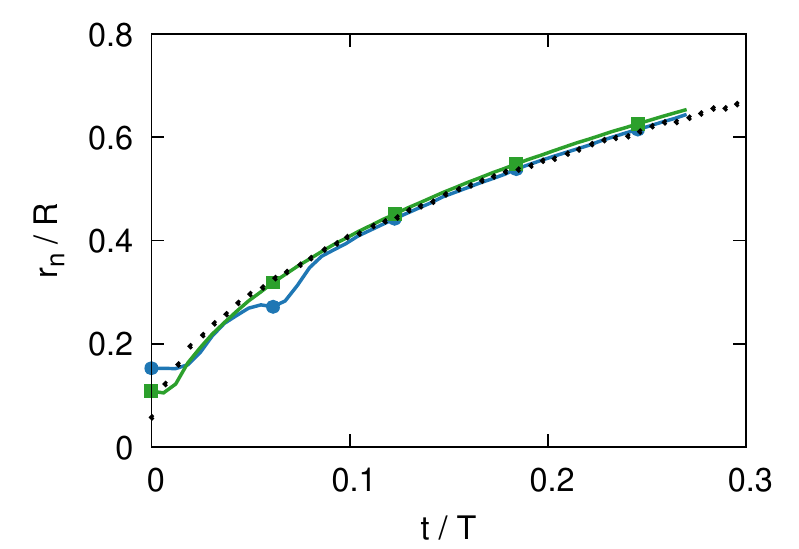}

  \caption{Evolution of the coalescence neck radius
  relative to the bubble radius depending on the resolution
    $R/h=38.4$~\key{011} and $76.8$~\key{032}
    produced by Aphros
    compared to experiment~\cite{thoroddsen2005} (dots).
  }
  \label{f:coalthorrn}
\end{figure}

\revvg{
We have also considered another experimental study
of bubble coalescence~\cite{thoroddsen2005}.
Initially, both bubbles are positioned along the~$z$-axis
and have elliptical shapes.
One bubble with semi-axes $R_x=R_y=1.08\,R$ and $R_z=1.0\,R$
is positioned above the other bubble
with semi-axes $R_x=R_y=0.92\,R$ and $R_z=0.97\,R$,
where~$R$ is their characteristic size.
The Ohnesorge number $\text{Oh} = \frac{\mu_1}{\sqrt{\rho_1R \sigma}}$
is set to $\text{Oh}=0.007$.
The experimental images overlaid with the results of the
simulation are shown in Figure~\ref{f:coalthor}
and the evolution of coalescence neck is presented in Figure~\ref{f:coalthorrn}.
In this case, our results agree with the experimental data
without scaling of time.
}

\begin{figure}
  \centering
  \def\coalvspace{\vspace{0.5mm}}
  \newcommand{\p}[1]{%
    \hfill%
    {\includegraphics[width=0.245\columnwidth]{coalwall/viewlightblue/a_#1.png}}%
    \hfill%
  }

  \p{00}%
  \p{01}%
  \p{02}%
  \p{03}

  \coalvspace

  \p{04}%
  \p{05}%
  \p{06}%
  \p{07}

  \coalvspace

  \p{08}%
  \p{09}%
  \p{10}%
  \p{11}

  \coalvspace

  \p{12}%
  \p{13}%
  \p{14}%
  \p{15}

  \coalvspace

  \p{16}%
  \p{17}%
  \p{18}%
  \p{19}

  \coalvspace

  \p{20}%
  \p{21}%
  \p{22}%
  \p{23}

  \caption{Snapshots from experiment~\cite{soto2018}
  overlaid by the projections of the shapes from the simulation
  by Aphros with resolution $R/h=76.8$ (blue lines)
  at times $t/T=$
  0, 0.13, 0.26, 0.4, 0.53, 0.66, 0.79, 0.92, 1.1, 1.2, 1.3,
  1.4, 1.6, 1.7, 1.8, 2.1, 2.4, 2.6, 2.9, 3.2, 3.7, 4.2, 4.7 and 5.3.
  The time values for the experimental images are multiplied by 1.2.
  }
  \label{f:coalwall}
\end{figure}

\begin{figure*}
  \centering
  \includegraphics{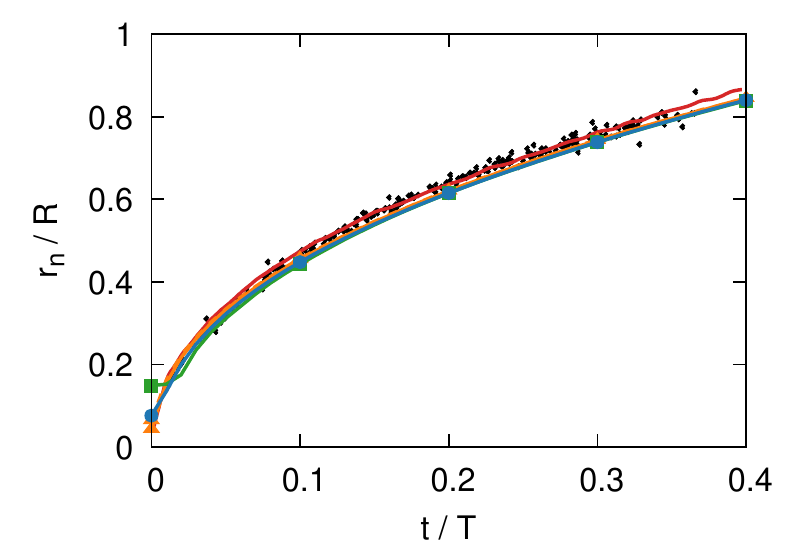}%
  \includegraphics{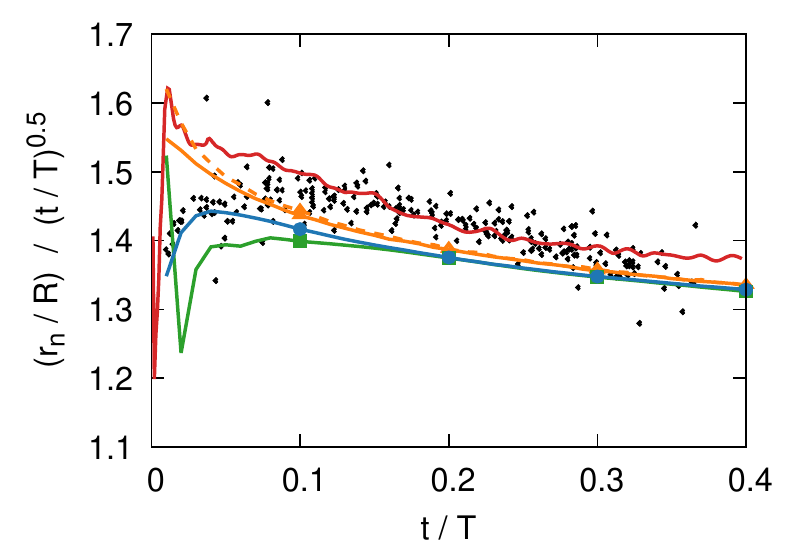}

  \caption{Evolution of the coalescence neck radius 
  relative to the bubble radius (left)
    and the coalescence neck radius divided by $\sqrt{t}$ (right).
    Gerris in the axisymmetric formulation with adaptive mesh refinement
    with the equivalent resolution
    $R/h=307.2$~\key{123} and $R/h=614.4$~\key{023},
    Basilisk with proposed method for curvature at $R/h=38.4$~\key{032}
    and Aphros at $R/h=153.6$~\key{011}.
    Experiment~\cite{soto2018} (dots)
    and boundary integral~\cite{soto2018}~\key{040} without scaling of time.
  }
  \label{f:rnge}
\end{figure*}

\subsection{Derivatives of particle positions}
\label{s:partderiv}

The algorithm in Section~\ref{s:meth2d}
requires derivatives of particle positions~(\ref{e:partpos})
with respect to angles~$\phi$ and $\theta$
which are given as
\begin{equation}
  \label{e:partderiv_ph}
  \frac{\pd\mb{x}_i}{\pd \phi} =
  \begin{cases}
   \sum\limits_{j=1}^{i-c} h_p
    \mb{e}\big(\phi + (j-\frac{1}{2})\,\theta+\frac{\pi}{2}\big) &\qquad i> c,
    \\
  \mb{0} &\qquad i=c,
    \\
  -\sum\limits_{j=1}^{c-i} h_p
    \mb{e}\big(\phi - (j-\frac{1}{2})\,\theta+\frac{\pi}{2}\big) &\qquad i<c,
  \end{cases}
\end{equation}
\begin{equation}
  \label{e:partderiv_th}
  \frac{\pd\mb{x}_i}{\pd \theta} =
  \begin{cases}
   \sum\limits_{j=1}^{i-c} h_p
    \big(j-\frac{1}{2}\big)
    \,\mb{e}\big(\phi + (j-\frac{1}{2})\,\theta+\frac{\pi}{2}\big) 
    &\qquad i> c,
    \\
  \mb{0} &\qquad i=c,
    \\
  \sum\limits_{j=1}^{c-i} h_p
    \big(j-\frac{1}{2}\big)
    \,\mb{e}\big(\phi - (j-\frac{1}{2})\,\theta+\frac{\pi}{2}\big) 
    &\qquad i<c,
  \end{cases}
\end{equation}

The following recurrence relations are useful
for the implementation:
\begin{equation}
  \label{e:rec_x}
  \mb{x}_i =
  \begin{cases}
    \mb{x}_{i-1} +
    h_p \mb{e}\big(\phi + (i-c-\frac{1}{2})\,\theta\big) 
    &\quad i> c,
    \\
  \mb{p} &\quad i=c,
    \\
    \mb{x}_{i+1}
    -h_p \mb{e}\big(\phi - (c-i-\frac{1}{2})\,\theta\big) 
    &\quad i<c,
  \end{cases}
\end{equation}
\begin{equation}
  \label{e:rec_ph}
  \frac{\pd\mb{x}_i}{\pd \phi} =
  \begin{cases}
    \frac{\pd\mb{x}_{i-1}}{\pd \phi} +
    h_p \mb{e}\big(\phi + (i-c-\frac{1}{2})\,\theta+\frac{\pi}{2}\big) 
    &\quad i> c,
    \\
  \mb{0} &\quad i=c,
    \\
    \frac{\pd\mb{x}_{i+1}}{\pd \phi}
    -h_p \mb{e}\big(\phi - (c-i-\frac{1}{2})\,\theta+\frac{\pi}{2}\big) 
    &\quad i<c,
  \end{cases}
\end{equation}
\begin{equation}
  \label{e:rec_th}
  \frac{\pd\mb{x}_i}{\pd \theta} =
  \begin{cases}
    \frac{\pd\mb{x}_{i-1}}{\pd \theta} +
    h_p (i-c-\frac{1}{2})
    \,\mb{e}\big(\phi + (i-c-\frac{1}{2})\,\theta+\frac{\pi}{2}\big) 
    &\quad i> c,
    \\
  \mb{0} &\quad i=c,
    \\
    \frac{\pd\mb{x}_{i+1}}{\pd \theta} +
    h_p (c-i-\frac{1}{2})
    \,\mb{e}\big(\phi - (c-i-\frac{1}{2})\,\theta+\frac{\pi}{2}\big) 
    &\quad i<c,
  \end{cases}
\end{equation}

\subsection{Proof of convergence of iterations}
\label{s:proof}

We show that iterations in Section~\ref{s:meth2d}
can be formulated as steps of the gradient descent
minimizing an energy function.
Given the initial conditions,
the central particle~$\mb{x}_c$ already belongs to a line segment.
Therefore, the corresponding force is zero $\mb{f}^m_c=0$,
and Step~1 trivializes such that $\mb{p}^m=\mb{p}^0$,
$\mb{X}^*=\mb{X}^m$ and $\mb{F}^*=\mb{F}^m$.
In the following, the argument $\mb{p}$ is omitted.
We define the energy in terms of the remaining parameters $\phi$ and $\theta$
\begin{equation}
  E(\phi,\theta)
  =
  \frac{1}{2}
  \sum_{i=1}^N \min_{\mb{y}\in L} |\mb{y} - \mb{x}_i(\phi,\theta)|^2.
\end{equation}
The energy can be expressed as a superposition
\begin{equation}
  \label{e:Esuper}
  E(\phi,\theta) = 
  \hat{E}(\phi, \theta, \mb{Y}_\text{min}(\phi,\theta))
\end{equation}
of two functions
\begin{gather}
  \hat{E}(\phi,\theta,\mb{Y})
  =
  \frac{1}{2}
  \sum_{i=1}^N |\mb{y}_i - \mb{x}_i(\phi,\theta)|^2
  \\
  \label{e:ymin}
  \mb{Y}_\text{min}(\phi, \theta)
  = 
  [\argmin_{\mathbf{y}\in L}|\mb{y} - \mb{x}_i(\phi, \theta)|
  ,\;i=1,\dots,N].
\end{gather}
Derivatives of $\hat{E}$ read
\begin{gather*}
  \frac{\pd\hat{E}}{\pd \phi}(\phi,\theta,\mb{Y})
  =
  -(\mb{Y}-\mb{X}(\phi, \theta))\cdot \frac{\pd\mb{X}}{\pd\phi}(\phi, \theta)
  \\
  \frac{\pd\hat{E}}{\pd \theta}(\phi,\theta,\mb{Y})
  =
  -(\mb{Y}-\mb{X}(\phi, \theta))\cdot \frac{\pd\mb{X}}{\pd\theta}(\phi, \theta).
\end{gather*}
In this notation, the force defined by~(\ref{e:force}) and (\ref{e:forcevect})
transforms to
\begin{equation}
  \mb{F}(\phi, \theta) =
  \eta\,\big(\mb{Y}_\text{min}(\phi, \theta)-\mb{X}(\phi, \theta)\big).
\end{equation}
Taking into account
\begin{gather*}
  \mb{F}^* \cdot \mb{D}_\phi
  =
  -\eta \frac{\pd\hat{E}}{\pd \phi}(\phi^m,\theta^m,\mb{Y}^m)
  \\
  \mb{F}^* \cdot \mb{D}_\theta
  =
  -\eta 
  \Big[
  \frac{\pd\hat{E}}{\pd \theta}(\phi^{m+1},\theta^m,\mb{Y}^m)
  -
  \big(\mb{X}(\phi^{m+1}, \theta^m)-\mb{X}(\phi^{m}, \theta^m)\big)
  \Big]
  \,,
\end{gather*}
corrections in Step 2 and Step 3 can be expressed as
\begin{gather}
  \label{e:step2grad}
  \phi^{m+1} = \phi^m
  -\frac{\eta}{\|\mb{D}_\phi\|^2}
  \frac{\pd \hat E}{\pd \phi}
  (\phi^m, \theta^m, \mb{Y}^m)
  \\
  \label{e:step3grad}
  \theta^{m+1} = \theta^m
  -\frac{\eta}{\|\mb{D}_\theta\|^2}
  \frac{\pd \hat E}{\pd \theta}
  (\phi^{m+1}, \theta^m, \mb{Y}^m)\,-
  \\\notag
  -\frac{1-\eta}{\|\mb{D}_\theta\|^2}
  \big(\mb{X}(\phi^{m+1}, \theta^m)-\mb{X}(\phi^{m}, \theta^m)\big)
  \cdot \mb{D}_\theta\,,
\end{gather}
where 
$\mb{F}^*=\mb{F}(\phi^m, \theta^m)$,
$\mb{D}_\phi=\frac{\pd\mb{X}}{\pd\phi}(\phi^m, \theta^m)$,
$\mb{D}_\theta=\frac{\pd\mb{X}}{\pd\theta}(\phi^{m+1}, \theta^m)$
and
$\mb{Y}^m=\mb{Y}_\text{min}(\phi^m, \theta^m)$.
The second correction can be rewritten
\begin{gather}
  \label{e:step3grade}
  \theta^{m+1} = \theta^m
  -\frac{\eta}{\|\mb{D}_\phi\|^2}
  \frac{\pd \hat E}{\pd \theta}
  (\phi^{m+1}, \theta^m, \mb{Y}^m)\,+
  \\\notag
  +\frac{(1-\eta)\eta \,\tilde{\mb{D}}_\phi \cdot \mb{D}_\theta }
  {\|\mb{D}_\phi\|^2\|\mb{D}_\theta\|^2}
  \frac{\pd \hat E}{\pd \phi}
  (\phi^m, \theta^m, \mb{Y}^m)
\end{gather}
in terms of the finite difference
\begin{equation}
  \label{e:tildedphi}
  \tilde{\mb{D}}_\phi =
  \frac{\mb{X}(\phi^{m+1}, \theta^m)-\mb{X}(\phi^{m}, \theta^m)}
  {\phi^{m+1} - \phi^{m}}.
\end{equation}
Function $\hat E(\phi,\theta,\mb{Y})$ is smooth on a compact set
$[0,2\pi] \times [0,2\pi] \times L^N$,
and therefore its gradient is Lipschitz continuous.
In particular, for a constant $\lambda>0$
\begin{equation}
  \label{e:lips}
  \begin{split}
    \Big|\frac{\pd \hat E}{\pd \phi}(\phi+\Delta\phi,\theta,\mb{Y})-
    \frac{\pd \hat E}{\pd \phi}(\phi,\theta,\mb{Y})\Big| \leq
    \lambda\,\Delta\phi,\\
    \Big|\frac{\pd \hat E}{\pd \theta}(\phi,\theta+\Delta\theta,\mb{Y})-
    \frac{\pd \hat E}{\pd \theta}(\phi,\theta,\mb{Y})\Big| \leq
    \lambda\,\Delta\theta.
  \end{split}
\end{equation}
This implies that
\begin{equation}
  \big| \Delta E - (E_\phi\Delta\phi+E_\theta\Delta\theta) \big| 
  \leq \lambda (\Delta\phi^2 +\Delta\theta^2),
\end{equation}
where
$\Delta\phi=\phi^{m+1}-\phi^m$, $\Delta\theta=\theta^{m+1}-\theta^m$,
$\Delta E=\hat{E}(\phi^{m+1},\theta^{m+1},\mb{Y}^m)-\hat{E}(\phi^{m},\theta^m,\mb{Y}^m)$,
$E_\phi=\frac{\pd \hat E}{\pd \phi}(\phi^{m}, \theta^m, \mb{Y}^m)$ and
$E_\theta=\frac{\pd \hat E}{\pd \theta}(\phi^{m+1}, \theta^m, \mb{Y}^m)$.
Therefore, the change of the energy after both corrections is bounded as
\begin{equation}
\Delta E\leq E_\phi\Delta\phi + E_\theta\Delta\theta+
  \lambda (\Delta\phi^2 +\Delta\theta^2).
\end{equation}
We show that for a sufficiently small~$\eta>0$,
\begin{equation}
E_\phi\Delta\phi + E_\theta\Delta\theta+
  \lambda (\Delta\phi^2 +\Delta\theta^2) \leq 0,
\end{equation}
which, given (\ref{e:step2grad}) and (\ref{e:step3grade}), is equivalent to
\begin{equation}
  -\mb{v}^T\mb{A}\mb{v} + \lambda\eta\,|\mb{A}\mb{v}|^2 \leq 0,
\end{equation}
where
\begin{equation}
  \mb{A}=
    \begin{bmatrix}
      \frac{1}{\|\mb{D}_\phi\|^2} & 0 \\
  -\frac{(1-\eta) \,\tilde{\mb{D}}_\phi \cdot \mb{D}_\theta}
  {\|\mb{D}_\phi\|^2\|\mb{D}_\theta\|^2}
      & \frac{1}{\|\mb{D}_\theta\|^2} &
    \end{bmatrix}
  \quad
  \mb{v} = 
    \begin{bmatrix} E_\phi \\ E_\theta \end{bmatrix}.
\end{equation}
This follows from a stronger inequality
\begin{equation}
  -\mb{v}^T\mb{A}\mb{v} + \lambda\eta\,\|\mb{A}\|^2|\mb{v}|^2 \leq 0
\end{equation}
or, in terms of a symmetric matrix,
\begin{equation}
  \mb{v}^T\Big(\frac{\mb{A}+\mb{A}^T}{2}-\lambda\eta\,\|\mb{A}\|^2\mb{I}\Big)\mb{v}
  \geq 0,
\end{equation}
which states that the matrix is positive semi-definite.
As seen from (\ref{e:partderiv_ph}-\ref{e:partderiv_th}),
derivatives $\mb{D}_\phi$ and $\mb{D}_\theta$
are uniformly bounded away from zero and infinity.
Therefore, it is sufficient to show that matrix
$\frac{\mb{A}+\mb{A}^T}{2}$
is positive definite in the limiting case $\eta\rightarrow 0$
\begin{equation}
  \frac{\mb{A}+\mb{A}^T}{2}=
    \begin{bmatrix}
      \frac{1}{\|\mb{D}_\phi\|^2} & 
      -\frac{1}{2}\frac{{\mb{D}}_\phi \cdot \mb{D}_\theta}
  {\|\mb{D}_\phi\|^2\|\mb{D}_\theta\|^2}
      \\
      -\frac{1}{2}\frac{{\mb{D}}_\phi \cdot \mb{D}_\theta}
  {\|\mb{D}_\phi\|^2\|\mb{D}_\theta\|^2}
      & \frac{1}{\|\mb{D}_\theta\|^2} &
    \end{bmatrix}
\end{equation}
as the finite difference $\tilde{\mb{D}}_\phi$ from~(\ref{e:tildedphi}) 
uniformly converges to $\mb{D}_\phi$.
The above is equivalent to
\begin{equation}
  \frac{ \,({\mb{D}}_\phi \cdot \mb{D}_\theta)^2}
  {\,\|\mb{D}_\phi\|^2\|\mb{D}_\theta\|^2} < 4,
\end{equation}
which always holds for a scalar product.
This shows that corrections of parameters~$\phi$ and $\theta$ reduce the energy
\begin{equation}
\hat{E}(\phi^{m+1},\theta^{m+1},\mb{Y}^m)\leq\hat{E}(\phi^{m},\theta^m,\mb{Y}^m).
\end{equation}
Finally, definition (\ref{e:ymin}) implies that
\begin{equation}
  \hat E(\phi^{m+1},\theta^{m+1},\mb{Y}_\text{min}(\phi^{m+1},\theta^{m+1}))
  \leq
  \hat E(\phi^{m+1},\theta^{m+1},\mb{Y}^m).
\end{equation}
Combining the last two inequalities,
we arrive at
\begin{equation}
  \hat E(\phi^{m+1},\theta^{m+1},\mb{Y}^{m+1})\leq
  \hat E(\phi^{m},\theta^{m},\mb{Y}^{m})
\end{equation}
or, in the original notation (\ref{e:Esuper}),
\begin{equation}
  E(\phi^{m+1},\theta^{m+1})\leq
  E(\phi^{m},\theta^{m}).
\end{equation}
Together with $E \geq 0$, this guarantees convergence of iterations.


\section*{\refname}
\begingroup
\renewcommand{\section}[2]{}%
\bibliography{main}
\endgroup

\end{document}